\documentclass[amsfonts, amssymb, amsmath, aps, prd, 10pt, pra, twocolumn, superscriptaddress, english, floatfix, longbibliography]{revtex4-2}

\usepackage{soul}
\usepackage{braket}
\usepackage{amsfonts} 
\usepackage{graphicx}
\usepackage{dcolumn}
\usepackage{epsfig}
\usepackage{bm}
\usepackage{array}
\usepackage{hyperref}
\hypersetup{
colorlinks=true,
citecolor=blue,
linkcolor=blue,
urlcolor=blue,
}
\usepackage{xcolor}			
\usepackage{soul} 
\usepackage{algorithm}
\usepackage{algpseudocode}

\usepackage{bbm}
\usepackage[english]{babel}
\usepackage{setspace}
\usepackage{wasysym}
\usepackage{adjustbox}
\usepackage{framed}
\usepackage{latexsym} 
\usepackage{multirow} 
\usepackage{booktabs} 
\usepackage{calc} 
\usepackage[normalem]{ulem}
\usepackage[inkscapeformat=png]{svg}
\usepackage{tikz}

\DeclareMathAlphabet{\pazocal}{OMS}{zplm}{m}{n}	

\definecolor{DeepSkyBlue}{RGB}{0,104,139}
\colorlet{MySky}{white!40!blue}
\colorlet{MyViolet}{red!45!blue}
\colorlet{MyBlue}{black!40!blue}
\colorlet{MyRed}{black!40!red}
\colorlet{MyOrange}{red!70!yellow}
\colorlet{MyGreen}{black!60!green}
\colorlet{MyBrown}{black!70!brown}
\colorlet{MyGray}{black!60!white}

\newcommand{\be}{\begin{equation}}
\newcommand{\ee}{\end{equation}}

\newcommand{\beq}{\begin{eqnarray}}
\newcommand{\eeq}{\end{eqnarray}}

\begin{document}

\title{Detecting quasi-degenerate ground states in 
 topological models via variational quantum eigensolver}

\author{Carola Ciaramelletti}
\email[correspondence at: ]{carola.ciaramelletti@graduate.univaq.it}
\affiliation{Dipartimento di Ingegneria e Scienze dell'Informazione e Matematica, Universit\`a dell'Aquila, via Vetoio,
I-67010 Coppito-L'Aquila, Italy.}

\author{Martin Beseda}
\affiliation{Dipartimento di Ingegneria e Scienze dell'Informazione e Matematica, Universit\`a dell'Aquila, via Vetoio,
I-67010 Coppito-L'Aquila, Italy.}
\affiliation{IT4Innovations, VSB - Technical University of Ostrava, 17. listopadu 2172/15, Ostrava, 708 00, Czech Republic.}

\author{Mirko Consiglio}
\affiliation{Department of Physics, University of Malta, Msida MSD 2080, Malta.}

\author{Luca Lepori}
\affiliation{Dipartimento di Scienze Matematiche, Fisiche e Informatiche, Universit\`a di Parma, Parco Area delle Scienze, 53/A, I-43124 Parma, Italy.}
\affiliation{INFN, Gruppo Collegato di Parma, Parco Area delle Scienze 7/A, 43124, Parma, Italy.}

\author{Tony J. G. Apollaro}
\affiliation{Department of Physics, University of Malta, Msida MSD 2080, Malta.}

\author{Simone Paganelli}
\affiliation{Dipartimento di Scienze Fisiche e Chimiche, Universit\`a dell'Aquila, via Vetoio,
I-67010 Coppito-L'Aquila, Italy.}

\begin{abstract}
 We study the exact ground states of the Su--Schrieffer--Heeger open chain and of the Kitaev open chain, using the Variational Quantum Eigensolver (VQE) algorithm. These models host symmetry-protected topological phases, characterized by edge modes with vanishing single-particle energy in the thermodynamic limit. The same fact prevents the standard VQE algorithm from converging to the correct ground state for finite chains, since it is quasi-degenerate in energy with other many-body states. Notably, this quasi-degeneracy cannot be removed by small perturbations, as in typical spin systems. We address this issue by imposing appropriate constraints on the VQE evolution and constructing appropriate variational circuits, to restrict the probed portion of the Hilbert space along the same evolution. These constraints stem from both general properties of the topological phases and of the studied Hamiltonians. In this way, the improved VQE algorithm achieves an accurate convergence to the exact ground states in each phase. The present approach promises large applicability, also to realistic systems with different topologies and/or not easily removable degeneracies, thanks to the very high fidelity achievable also on systems with a relatively high number of qubits.
\end{abstract}

\bigskip

\maketitle

\section{Introduction}

Quantum computing is a rapidly evolving field that explores the potential of harnessing the principles of quantum mechanics for computational speed-up. At its core, quantum computing relies on the fundamental principles of quantum mechanics, such as superposition, entanglement, and interference~\cite{nielsen2010quantum}. Currently, quantum computation exists in the Noisy Intermediate-Scale Quantum (NISQ) era \cite{Preskill2018}, which represents a crucial phase in the development of quantum technologies, characterized by the control over single quantum systems, however critically subject to errors and noise. Nonetheless, several claims have been made about quantum calculations performed beyond the reach of classical computers~\cite{QuantumSupremacy,king2024computationalsupremacyquantumsimulation,YM2024}. 

Since current NISQ devices tend to suffer from rather strong decoherence, a great deal of research effort is focused on the field of hybrid quantum--classical algorithms, where a quantum computer is tasked with evaluating quantum expectation values embedded in a classical optimization framework. One family of such algorithms is the variational quantum algorithms, aimed at addressing problems in quantum chemistry, materials science, and many-body systems, among others~\cite{Peruzzo2014, fedorov2021vqe, Consiglio_2022,YM2024}.

Given the potential of quantum computing, it is essential to explore variational quantum algorithms' applications to different physical systems that can leverage quantum properties for enhanced computational capabilities. In this context, the study of topological systems in physics is becoming an area of growing interest in recent years due to their unique and intriguing properties \citep{bernevig2013topological}. These systems are mathematically characterized by certain topological invariants, i.e., nonlocal quantities that are insensitive to small changes in local parameters or thermodynamical quantities such as temperature or pressure. One of the most important features of many topologically non-trivial phases is the existence of topological surface states, that are robust against large sets of perturbations~\cite{RevModPhys.82.3045}. The existence of topological surface states is important for different reasons: first, in certain circumstances they provide a direct way to probe the bulk properties of a compound; second, they can have practical applications, such as in the development of new materials and structures, as well as for quantum simulation and information. For instance, topological systems have been shown to be suitable to implement qubits that are robust against decoherence~\cite{freedman2002topological, KITAEV20062, RevModPhys.80.1083}. Related to this fact, the study of topological systems could also play a central role in the development of new quantum algorithms, e.g., the topological properties of certain systems can be used to perform topological quantum error correction~\cite{Kitaev_2003, Dennis_2002}. 

Recently, quantum computation has been applied to investigate topological phases, for instance: a topological quantum phase transition has been parametrically crossed by the states generated on a quantum computer~\cite{PhysRevResearch.4.L022020} higher-order topological phases in high-dimensional lattices have been investigated~\cite{kohRealizationHigherorderTopological2024a}; using iterative quantum phase estimation; a classically optimized Variational Quantum Eigensolver (VQE) algorithm, supported by machine learning tools, has been utilized for measuring the string order parameters and to reveal a topological phase transition in certain exactly-solvable spin-$1/2$ systems~\cite{PhysRevResearch.5.043217}. Parallely, Kitaev spin models with genuine topological oder have been simulated in \cite{jahin2022} using a mapping on free fermions. For the latter large family of states, efficient approximations have been discussed in \cite{white2015}. Exploiting consequent techniques, a Fermi-Hubbard chain has been simulated  in \cite{niu2022} (using fermionic Gaussian matrix-product states  implementable on a trapped-ion quantum processor), and parity-defined sectors of the Kitaev chain sectors in \cite{rancic2022,sung2023}.  Finally, a simulation of a two-dimensional $Z_2$ lattice gauge theory (related to the Kitaev spin model mentioned above) on a shallow quantum circuit has been proposed later in \cite{burrello2022}.

The aim of this work is to study two models that host topologically non-trivial symmetry-protected phases, using variational quantum algorithms. Specifically, the VQE algorithm is employed in order to identify the topologically non-trivial ground states of the Su--Schrieffer--Heeger (SSH)~\cite{su1979} and Kitaev open chains~\cite{kitaev2001}.

The SSH chain is an instance of a one-dimensional topological insulator with spinless fermions hopping on a bipartite lattice, while the Kitaev chain is a prototype of topological superconductor, describing spinless fermions with nearest-neighbor superfluid pairing. These specific models are chosen since they are the simplest ones exhibiting non-trivial (symmetry-protected) topological behavior and are exactly solvable. In this respect, the exact solvability of the SSH and Kitaev chains makes them ideal playgrounds to test the convergence of the VQE, i.e. towards the ground states, in the different phases. Moreover, the presence of edge modes with vanishing energy in the thermodynamic limit makes the ground states of the topological phases quasi-degenerate at finite sizes with other many-body states; for this reason, the convergence of the prototypical VQE in these phases is generally poor, as also noticed in \cite{cugini2024}.
Notably, due to topology, this quasi-degeneracy cannot be removed by small perturbations, differently as, for instance, in typical spin systems.

In \cite{Sun2023}, the spin-1/2 Alternating Heisenberg Chain has been simulated with a shallow circuit, consisting of an initialisation layer and a variational layer. The degeneracy issue is solved however, the choice of the specific initialisation layer is subject to the a priori knowledge of the topological phase one aims to simulate. Degeneracies are not always
related with symmetries (see for instance \cite{RevModPhys.80.1083,fu2008,stern2012,clarke2013,kitaev1998,Kitaev_2003,bombin2010,wen2012}), 
such that they cannot be easily removed singling out definite symmetry sectors, as in \cite{rancic2022,sung2023}. The latter situation is conceptually similar to ~\cite{PhysRevResearch.5.043217}, where every known edge configuration is obtained separately, each time with a different circuit.
All these facts make the considered chains even more suitable to illustrate the power of our method.

Here, we solve this convergence issue, imposing appropriate constraints on the portion of the Hilbert space probed along the sequence of transformations performed by the VQE algorithm. These constraints stem from both general properties of the topological phases and of the studied Hamiltonians. In this way, the improved VQE achieves accurate convergence to the exact ground states in every phase. The present approach promises a large applicability, also on realistic systems with different types of topology, also thanks to the verified good behavior against enlarging the system sizes.
 
The paper is organized as follows: in section~\ref{models}, the SSH and the Kitaev model are briefly recalled. In section~\ref{VQEsim}, we present different VQE approaches to address the quasi-degeneracy of the ground states in the topological non-trivial phases of the aforementioned models. In section \ref{res}, we show the main results of the work. Finally, in section~\ref{concl}, we draw our conclusions. The paper is accompanied by Appendices~\ref{appconc},~\ref{adaptapp}, and \ref{warmapp},
where details and further results from the different VQE strategies are reported.

\section{The models}
\label{models}

In this section, we recall some basic features of the SSH and Kitaev chains, focusing on the properties of the symmetry-protected topological phases, which will be utilized to improve the VQE implementations.

\subsection{SSH chain}
\label{ssh}
The SSH model~\cite{su1979,asboth2016} describes spinless fermions hopping on a one-dimensional lattice with staggered hopping amplitudes \textit{v} and \textit{w} between two sublattices $A$ and $B$. Here, we consider an open chain with $N/2$ unit cells, each unit cell hosting two sites. Interactions between the particles are neglected, and the dynamics is described by the quadratic Hamiltonian
\begin{equation}
H = -v \sum_{j=1}^{N/2} (\hat{c}^{\dagger}_{j,A} \hat{c}_{j,B} + \text{h.c.}) - w \sum_{j=1}^{N/2 - 1} (\hat{c}^{\dagger}_{j,B} \hat{c}_{j+1,A} + \text{h.c.}).
\label{ssheq}
\end{equation}
The bulk properties can be studied first in the thermodynamic limit, where in the momentum space the Hamiltonian can be recast as
\begin{equation}
H=\sum_{k} \underline{c}_{k}^{\dag} H (k) \underline{c}_{k},
\end{equation}
where 
$
\underline{c}_{k}^{\dag}=
\begin{pmatrix}
c_{A,k}^{\dag}, & c_{B,k}^{\dag} \\
\end{pmatrix}
$
is the vector of creation operators acting on the $A$ and $B$ sublattices, $k$ lies inside the first Brillouin zone, i.e. $k \in [-\pi/a, \pi/a]$, where $a$ denotes the lattice spacing, and 
\begin{equation}
H (k) = \bm{d}\left(k\right) \cdot \bm{\sigma}_{k} \, ,
\label{Hmom}
\end{equation}
with
\begin{equation}
d_{x}(k)=v +w\cos(k); \quad d_{y}(k)=w\sin(k); 	\quad d_{z}(k)=0 \, .
\label{d}
\end{equation}
The Hamiltonian possesses a topology-protecting sublattice symmetry
\begin{equation}
 \sigma_z H (k) \sigma_z = -H (k) \, ,
\end{equation}
also reflected in a two-band symmetric spectrum, reading
\begin{equation}
E^{\pm}_{k}\left(\bm{d}_{k}\right) =\pm \sqrt{d_{x}^{2}+d_{y}^{2}} = \pm \sqrt{v^{2} + w^{2} + 2vw \cos(k)}~.
\label{dispersion}
\end{equation}
The SSH Hamiltonian belongs to the AIII-class in the ten-fold way classification~\cite{altland1997} for topological insulators and superconductors. 
Hereafter, we adopt the following parametrization for the parameters $v$ and $w$ 
 \begin{equation}
v = 1 - \delta, \qquad w = 1 + \delta. 
\end{equation}
Moreover, along with all the work, we will consider the half-filling regime $\nu = 1/2$, then the Fermi level sets in between the two bands in Eq. \eqref{dispersion}. In this condition, two different topological phases occur, depending on the sign of $\delta$: a trivial phase if $\delta<0$ ($v > w$) and a topologically non-trivial phase if $\delta>0$ ($v < w$).

In the topologically non-trivial phase, the system with open boundaries displays two symmetry-protected (i.e. against disorder or other spatial perturbations) modes, localized at the edges of the chain. These modes have zero single-particle energy in the thermodynamic limit \cite{Bernevig2013}. For a finite open chain, the two edge modes hybridize, giving rise to two single-particle states with energy $\pm \Delta E$. However, as $\Delta E$ vanishes exponentially with the number of sites, the hybridization amplitude also vanishes: $\Delta E \propto e^{-N/\xi}$, where $\xi$ is the localization length given by $1/\log(v/w)$~\cite{asboth2016}. 
In turn, these facts make the half-filling ground-state of the system  two-fold quasi-degenerate (quasi- at finite sizes) in energy, the two elements of this quasi-multiplet differing each-other by the presence of the hybridized mode described above. In particular, the ground-state has populated the hybrid state with energy $- \Delta E$. Also important for the following, this state displays two peaks at the edges in the square of the spatial wavefunction.

\subsection{Kitaev chain}
\label{kitaev}
The Kitaev chain~\cite{kitaev2001} describes spinless fermions with an on-site energy offset (chemical potential) and a nearest-neighbor superconductive pairing. 
The corresponding Hamiltonian reads
\begin{equation}
\begin{split}
H = & -\mu \sum_{n=1}^{N} c_{n}^{\dag}c_{n} - t \sum_{n=1}^{N-1}(c_{n+1}^{\dag} c_{n}+ \text{h.c.}) \\ 
& + \Delta \sum_{n=1}^{N-1}(c_{n}c_{n+1}+ \text{h.c}),
\end{split}
\label{Kc}
\end{equation}
where $\mu$ is the on-site potential, $t$ is the hopping term, and $\Delta$ is the pairing term creating or annihilating pairs of particles at neighboring sites. Experimental realizations of the model in Eq. \eqref{Kc} are claimed to have been performed in~\cite{kit1,kit2}. 

We cast the Hamiltonian in Eq.~\eqref{Kc} in momentum space, as
\begin{equation}
H(k)=
\begin{pmatrix}
-\mu-2t\cos(kd) & 2\Delta\sin(kd)\\
2\Delta \sin(kd) & \mu+2t\cos(kd) \\
\end{pmatrix} \, ,
\end{equation}
where now $\underline{c}_{k}=(c_{k}, c^{\dag}_{-k})^{T}$ is the Nambu vector. This Hamiltonian yields the excitation spectrum
\begin{equation}
E_{\pm}= \pm \sqrt{4 \Delta^{2} \sin^{2}(kd) + [\mu + 2t \cos(kd)]^{2}} \, ,
\end{equation}
where a topologically nontrivial phase is obtained for $|\mu/t|< 1/2$, hosting symmetry-protected edge modes, quoted Majorana modes \cite{kitaev2001} with vanishing energy. Via a very similar mechanism as for the SSH chain, this gives rise to a {\color{blue} two-fold} quasi-degeneracy of the many-body ground state at finite sizes, again due to the hybridization of the edge modes. We recall that here the ground state is characterized by all single-particle states with negative energy being populated, as for any Bogoliubov states \cite{ripka}. Again, the hybrid single-particle state with negative energy displays two peaks at the edges in the square of its spatial wavefunction.

\section{VQE SIMULATION APPROACHES} \label{VQEsim}

In this section, we report the study of the ground states of the two models introduced in the previous section by means of the VQE algorithm. With this aim in mind, we employ the Jordan--Wigner transformation \cite{giamarchi2003quantum}, mapping fermions to qubits (spin-$1/2$) and the SSH and the Kitaev chain, respectively, to an isotropic and anisotropic XY quantum spin-$1/2$ chain. 

\begin{figure*}[t]
\centering
\includegraphics[width=0.92\textwidth]{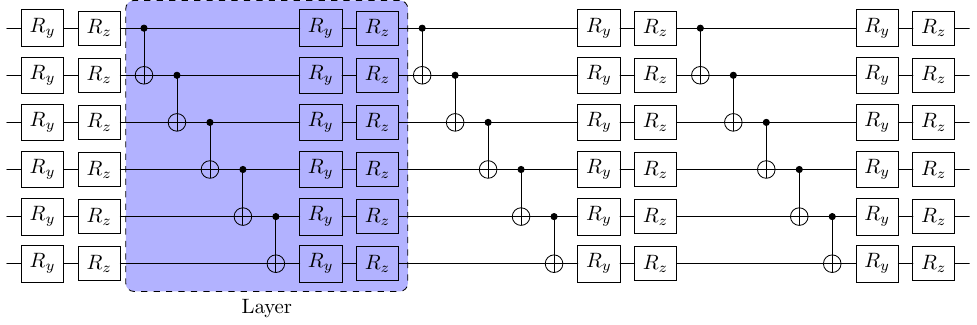}
\caption{Example of the employed hardware-efficient ansatz for $N=6$ qubits and $d=3$ repetitions of one layer. A single layer is characterized by a sequence of single-qubit parametric $R_y(\theta)$ and $R_z(\theta)$ gates, with a ladder of CNOT gates between nearest neighbors.}
\label{circuit}
\end{figure*}

In a first attempt to determine the ground state, we implement a standard VQE algorithm, minimizing the average energy with respect to the parameters of the ansatz. We use a hardware-efficient ansatz \cite{HardwareEff} that does not preserve the number of fermions, i.e., the total magnetization along the $z$-axis. This parametrized quantum circuit is a sequence of $d=3$ repetitions of one layer (in order to keep a sufficiently shallow circuit depth) of single-qubit rotations and CNOT gates, as shown in Fig.~\ref{circuit}.

\begin{figure}[t]
\centering
\includegraphics[width=0.49\textwidth]{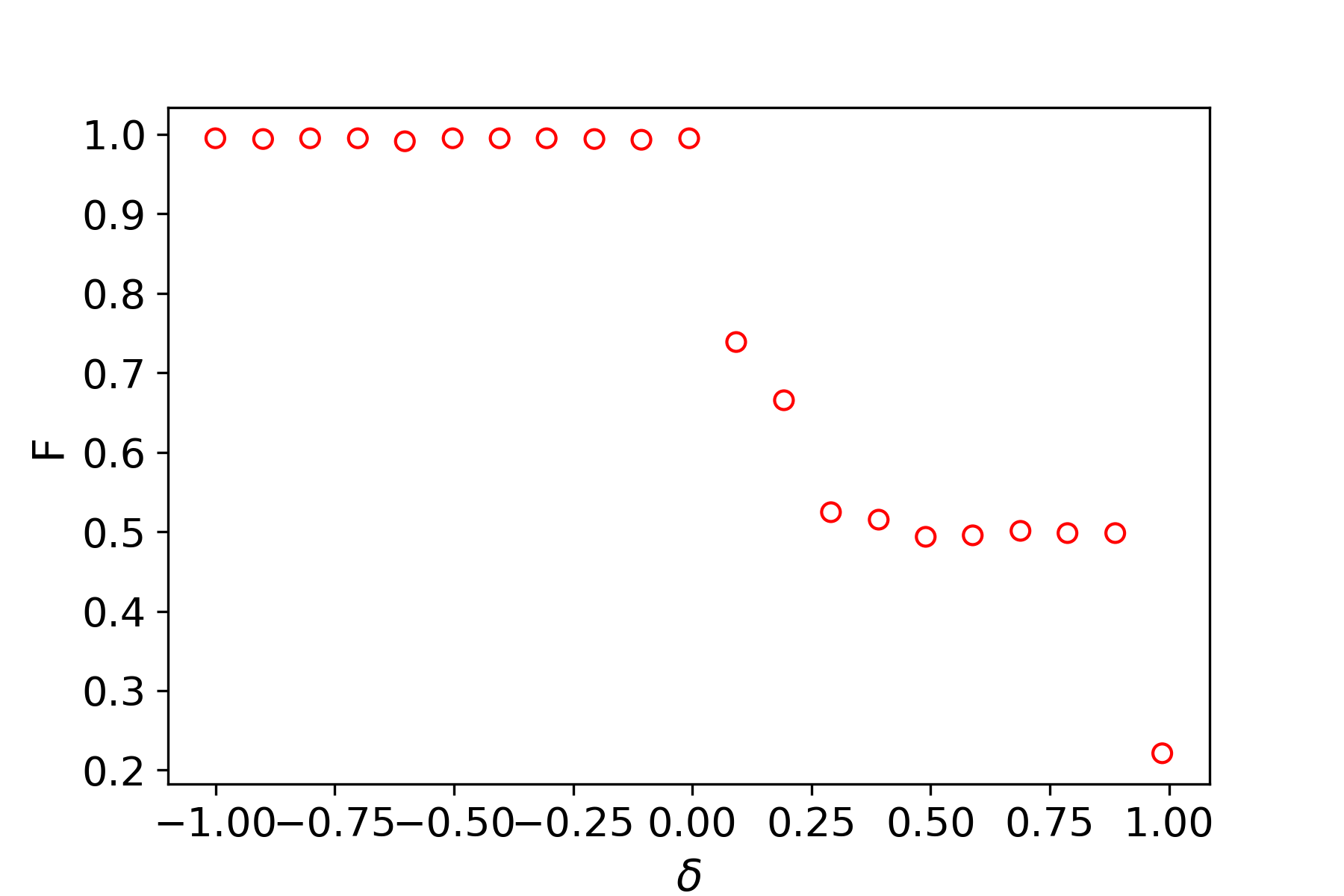}
\caption{Fidelity between the exact ground state and VQE state of the SSH chain with $N=6$ sites.
The VQE algorithm produces the correct ground state, as long as the system is in the trivial phase, while it fails in the non-trivial phase, where the fidelity drops.}
\label{conc_fid}
\end{figure}

\begin{figure}[t]
\centering
\includegraphics[width=0.49\textwidth]{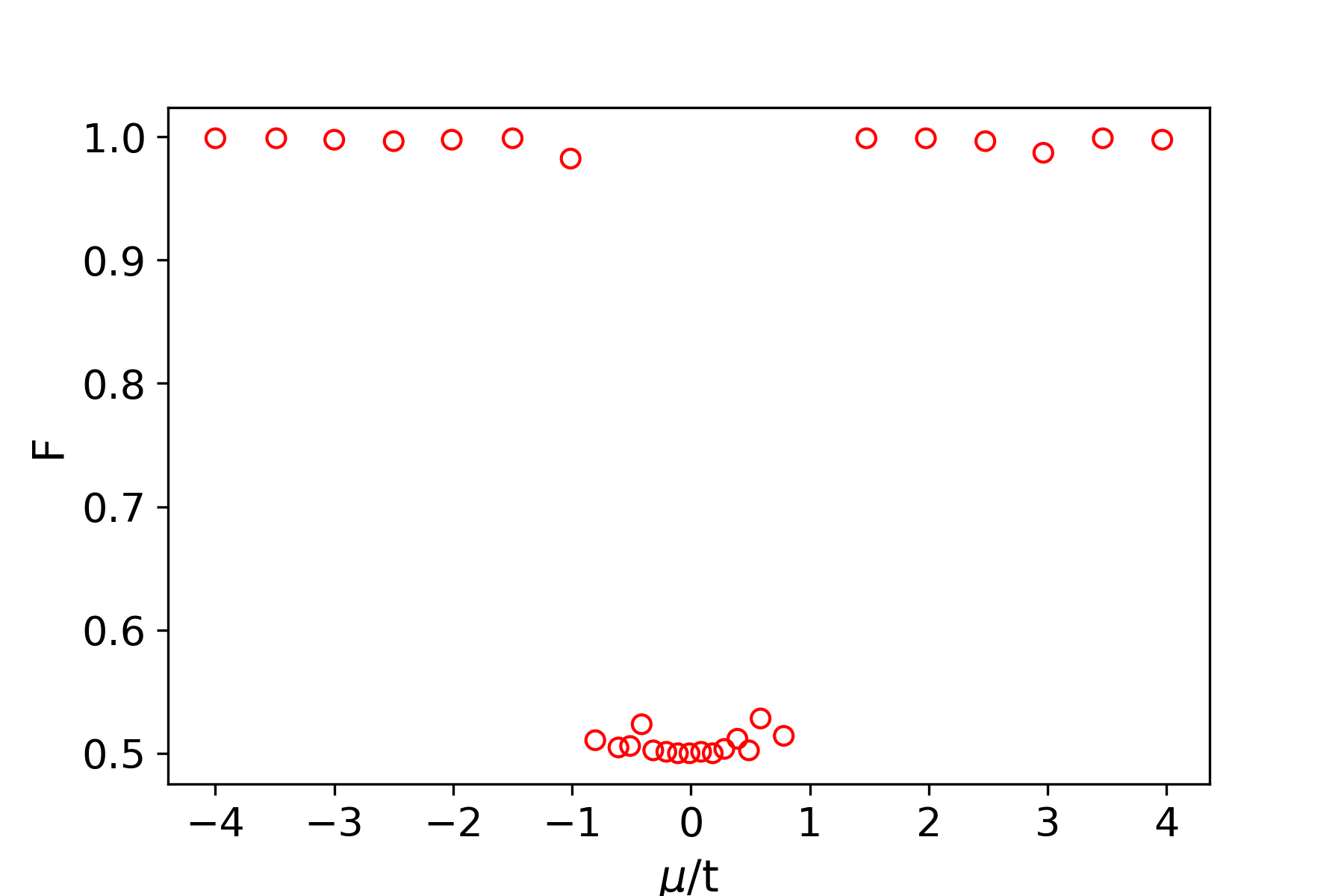}
\caption{Fidelity between the exact ground state for the Kitaev chain and the VQE states
as a function of $\mu/t$, for $N=6$ sites and $\Delta/t=1.2$. Again, the VQE algorithm produces the correct ground state, as long as the system is in the trivial phase, while it fails in the non-trivial phase, where the fidelity drops.}
\label{fidkit}
\end{figure}

To compare the state obtained by the VQE with the exact ground state, we compute their fidelity, defined as
\begin{equation}
F=\left|\braket{\psi_\text{VQE}|\psi_\text{Exact}}\right|^2.
\label{eq:Fid}
\end{equation}
We underline that the evaluation of the fidelity is not part of the algorithm and it is just computed at this stage as a comparison with the exact result. 
This is just a validation to proceed in a future step to  apply directly  the algorithm to more complex problems whose solution is not known.
The results for the SSH and Kitaev models are shown in Fig.~\ref{conc_fid} and Fig.~\ref{fidkit}, as functions of the parameter $\delta$ and $\mu$, respectively, for $N=6$ sites.
A sharp drop in the efficiency of the VQE algorithm appears evident when used to determine the ground state in the non-trivial phase. Indeed, since the many body ground state is quasi-degenerate in such a phase, the VQE ends up stuck in a subspace whose states have similar average energy. As a consequence, the algorithm gives a satisfactory ground state energy but fails in identifying the correct ground state. Being this a distinctive feature of phases with degenerate ground states, as the topological phase considered here, it is therefore necessary to find a strategy beyond the standard VQE approach.

The main idea is to exploit some general features of the ground states in the non-trivial phases and of the corresponding Hamiltonians to drive the minimization process in the degenerate subspace toward states closer to the exact ground state.

Different approaches can be adopted in order to account for these properties:
\begin{itemize}
 \item \textbf{Modified cost function}: the cost function in the VQE algorithm can be modified so that it would favor the states having the correct entanglement. The relevant properties we focus on are the amount of entanglement between two bulk sites and/or between the edges, detected by \emph{i}) the concurrence $C$ between them, and \emph{ii}) the purity $\mathcal{P}$ of two nearest-neighbor spins. Both these quantities, which drastically change across the two phases, are recalled in detail in the appendix \ref{appconc}.   
As explained in more detail in appendix \ref{adaptapp},  entanglement-related constraints are taken into account, together with the average energy in an adapted cost function of the form
\begin{equation}
\mathcal{C}(\bm{\theta})=E(\bm{\theta}) - \eta \, C\left(\bm{\theta})\right) + \tau \, \mathcal{P}\left(\bm{\theta}\right).
\label{cost_purity_0}
\end{equation} 

 \item \textbf{Warm-Start}: the initial state of the VQE can be chosen in such a way to be already sufficiently close to the state to be found. To improve the previous strategy, for instance, one can first implement a modified cost function approach described above, and then take the so-obtained state as the initial state for a new VQE where the extra constraints in the cost function are removed. Another option is to start from a trivial phase (where the VQE easily finds the correct ground state) close to a transition point with a topological phase, then change some Hamiltonian parameters by sufficiently small amounts, such as to move towards the second phase. At any step, the ground state is computed and used as the initial state for the next step. 
 \item \textbf{Problem-Inspired ansatz}: as a third approach, one could incorporate the knowledge of the models and their topological properties directly into the initial variational wave function, by constructing a `problem-inspired ansatz', that is a specifically designed ansatz considering some features (as from topology) of the Hamiltonian under investigation. From our results, this method, possibly employed together with the second one, appears a posteriori to be the most effective, both in terms of precision of the results and scalability.
\end{itemize}

All these strategies can help to improve the results of the VQE, nevertheless, each of them has drawbacks that need to be considered. In the first approach, changing the cost function seems, at first sight, quite straightforward and does not imply a change in the circuit setup. Nevertheless, since the main significant properties that we have identified as relevant are the entanglement properties of the system, any related quantity requires the evaluation of reduced density matrices. This task becomes operatively involved by increasing the size of the system.

The second approach can be employed by changing the Hamiltonian parameters by small amounts and without imposing extra constraints on the evolution along the quantum circuit. Then the constraints can be set as a control to exclude the output states that do not display the required properties. A real-amplitude ansatz~\cite{Qiskit} is used for the variational circuit, which consists of parametrized single-qubit $R_y$ gates and a ladder of \textsc{CNOT} gates as entangling gates. This ansatz often has fewer parameters than other ansatzes, and by limiting the parameter count, the optimization process can become more efficient and less prone to overfitting or getting stuck in local minima. Moreover, with a reduced parameter space, the optimization landscape can be simpler. This can lead to more accurate results and make optimization algorithms more effective. A drawback of this type of ansatz is that it is not scalable with the number of qubits required to describe the target state, as the barren plateau problem, i.e., the gradients of the cost functions become exponentially small making parameter training difficult \cite{larocca2024reviewbarrenplateausvariational}, is more prominent in this context. 
Moreover, the request for a small variation of the Hamiltonian parameters implies the need for a large number of runs of the quantum computer. These two approaches will be discussed in the appendix \ref{adaptapp} and \ref{warmapp}, respectively.

The last approach, being not hardware efficient, requires to be adapted to native gates of the quantum hardware and thus may produce a significant overhead of quantum gates. However, the problem-inspired ansatz appears as the most promising one in terms of scalability and results. 

Hereafter, we show the results for the two considered models, obtained via the problem-inspired ansatz and using the L-BFGS-B optimization algorithm~\cite{Nocedal2006}. The other strategies are also discussed in better detail in appendix \ref{adaptapp} and \ref{warmapp}.

\section{Main results from optimized hybrid strategy}
\label{res}
\begin{figure}[t]
\centering
\includegraphics[width=0.40\textwidth]{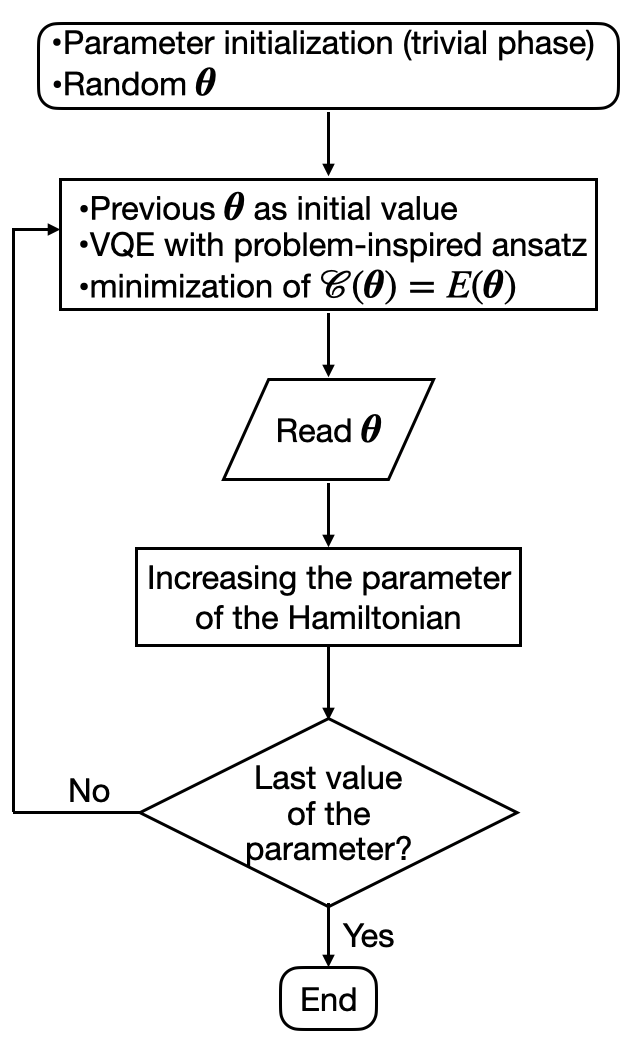}
\caption{Scheme of the problem-inspired ansatz  VQE  adopted.}
\label{diag4}
\end{figure}

\begin{figure*}
\centering
\includegraphics[width=0.9\textwidth]{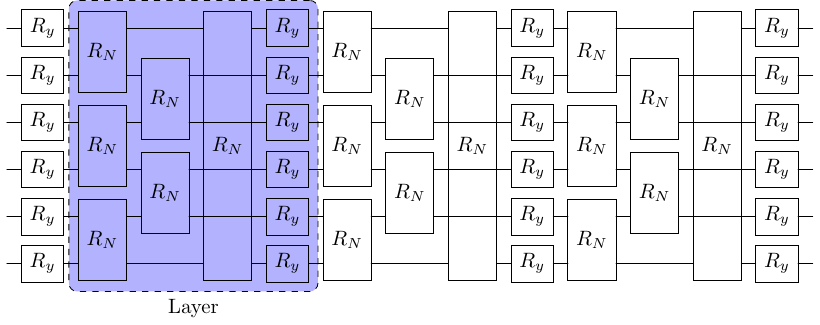}
\caption{Problem-Inspired circuit diagram for $N = 6$ qubits, used both for the SSH and for the Kitaev chains. The corresponding ansatz consists of a series of single-qubit parametric $R_y(\theta)$ gates and an alternating network of two-qubit parametric $R_N(\theta)$ gates, defined in Eq.~\eqref{eq:number_preserving}, that recreate the physics of the two models by generating correlations between neighboring and edge sites.}
\label{qulacs_new_Ansatz}
\end{figure*}

In this section, we adopt an optimized mix between the second and third strategies outlined above. In particular, we resort to a problem-inspired ansatz approach to incorporate the physics of the SSH model into the wavefunction represented by the quantum circuit. The scheme is depicted in Fig.\ref{diag4}.
The circuit includes rotation gates $R_y(\theta)$, which allow for the preparation of superpositions of the basis states, and the two-qubit gates 
\begin{equation} \label{eq:number_preserving}
 R_N(\theta) \equiv e^{i \frac{\theta}{2} (\sigma_x \otimes \sigma_y - \sigma_y \otimes \sigma_x)} \, , 
\end{equation}
effectively recreating the hopping elements in the SSH Hamiltonian. The latter gates are specifically chosen to generate correlations between neighboring sites. The same gates mimic the staggered nearest-neighbor interactions in the isotropic XY Hamiltonian, to which Eq. \eqref{Kc} is mapped via the Jordan-Wigner transformation. Additionally, they connect the first and the last qubit, inducing correlations between the edges as in a topologically non-trivial phase. Moreover, for each run, we slightly increase the parameter $\delta$ and use the state obtained in the previous run as a new initial state. A representative image of this circuit is shown in Fig.~\ref{qulacs_new_Ansatz}.

\begin{figure}[t]
\centering
\includegraphics[width=0.49\textwidth]{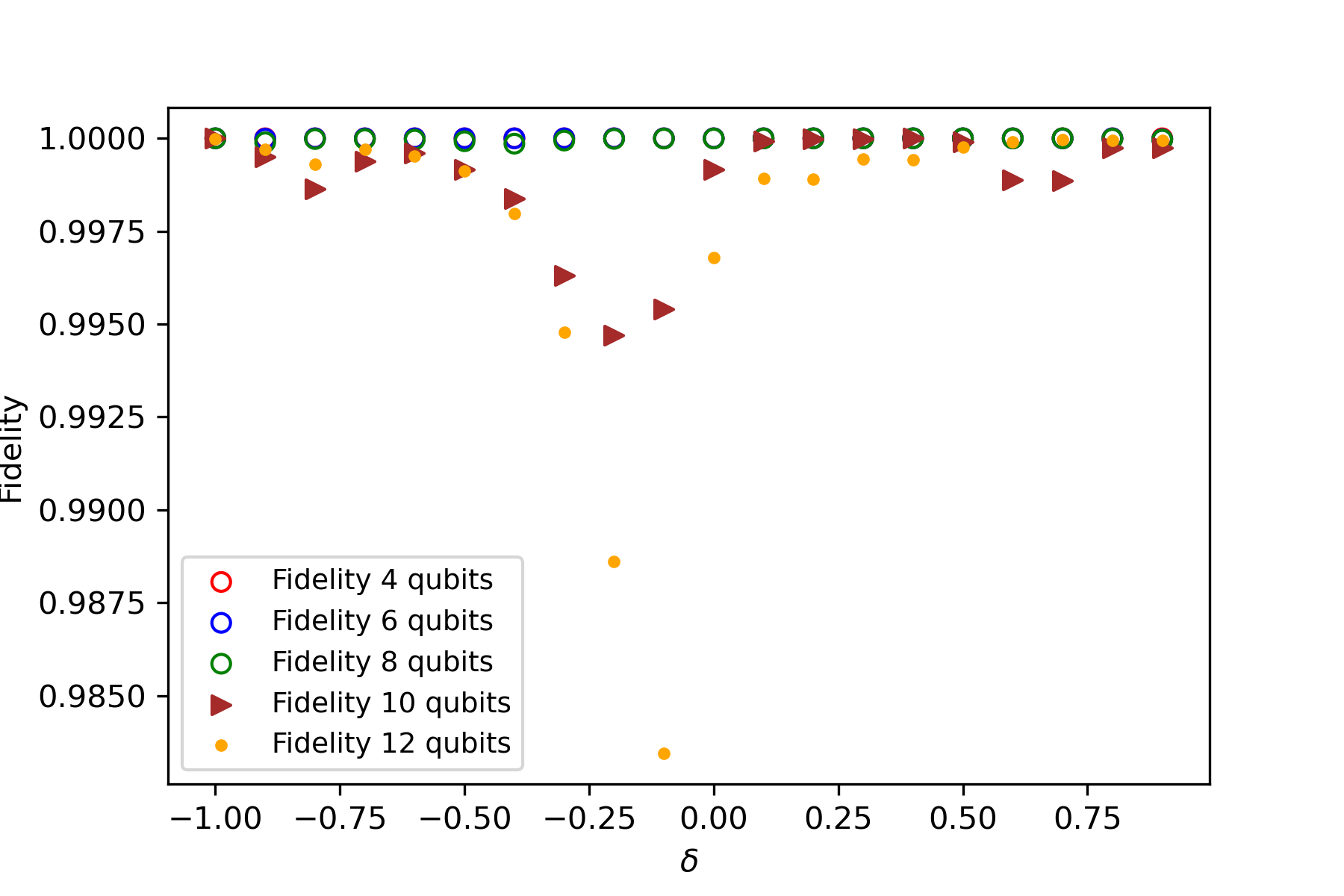}
\caption{Fidelity between the exact ground state and the VQE state for the SSH model. Results for $N=4,6,8,10,12$ qubits. High fidelity is achieved across the entire range of $\delta$. }
\label{ssh-statevector}
\end{figure}

The results of the state-vector simulations with this method are shown in Fig. \ref{ssh-statevector}, where we report the fidelity in Eq.~\eqref{eq:Fid} between the exact ground state and the VQE state, varying $\delta$ and for $N= 4, 6, 8, 10, 12$ sites. 
It can be observed that the convergence to the ground state is guaranteed across the entire range of $\delta$. It is also worth noticing that the present method allows this convergence without utilizing any properties of the topological system other than the Hamiltonian ones themselves. This fact avoids the requirement of precise knowledge not only of the properties of the searched ground state, but also of its topological quantum phase transition point. 
As a consequence, by evaluating the entanglement between two qubits of the obtained VQE solution, e.g., the first and last qubit, it is possible to determine whether, for the specified Hamiltonian parameters, the system is in the trivial or non-trivial topologically phase, thus determining also the topological quantum phase transition point.
Notably, in all the simulations, the number of layers scales logarithmically with the number of sites, specifically as $\lceil\log_{2}(n)\rceil+1$, making the present approach suitable for the investigation of larger systems.

\begin{figure}[h!]
    \centering
    \includegraphics[width=0.52\textwidth]{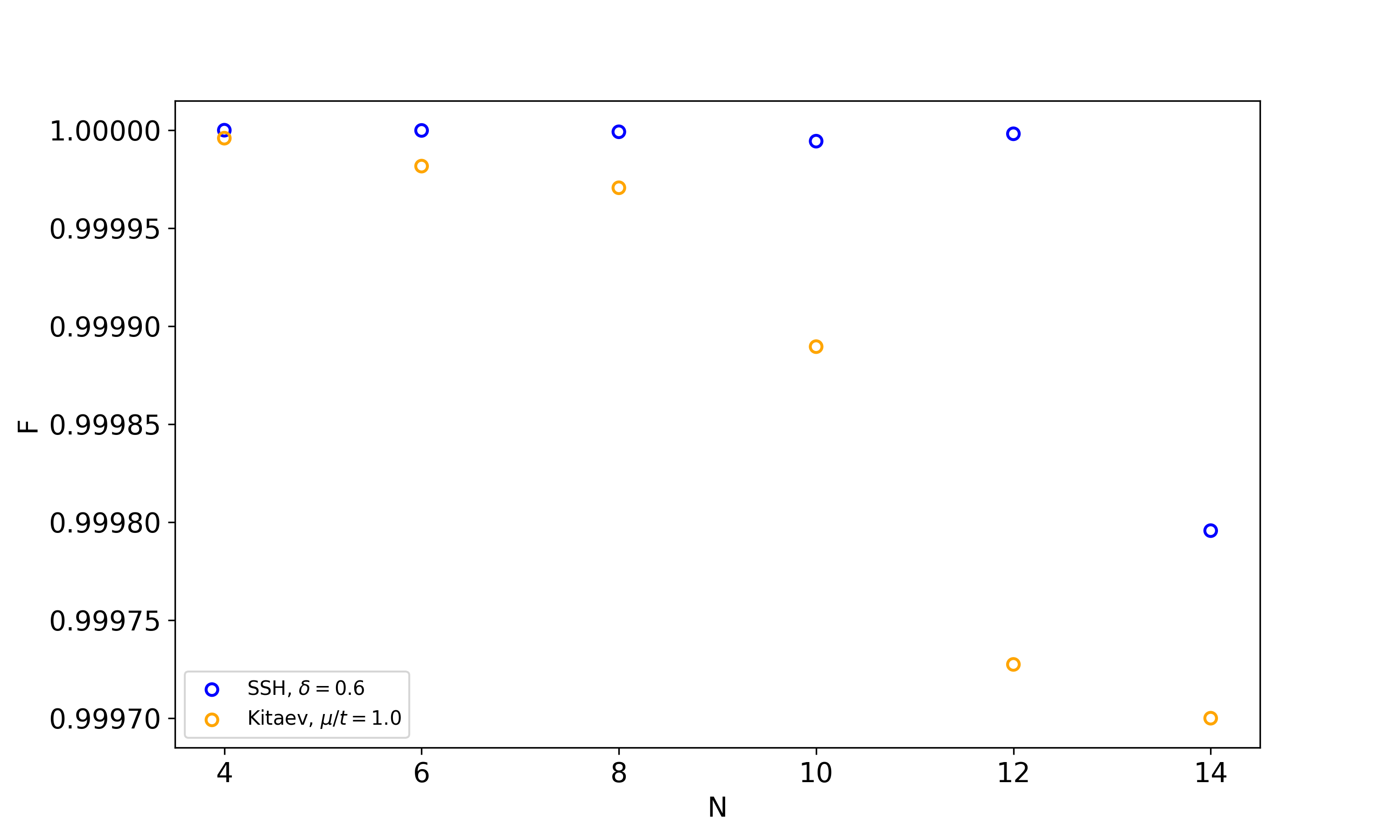}
    \caption{Fidelity as a function of system size \(N\) for the SSH model (\(\delta = 0.6\), blue circles) and Kitaev chain model (\(\mu/t = 1.0\), orange circles), both in their topologically non-trivial phases. The fidelity remains exceptionally high (\(> 0.999\)) for all tested system sizes, with a gradual and regular decrease as \(N\) increases.}
    \label{scaling}
\end{figure}

Also for the Kitaev chain, the problem-inspired approach produces more accurate results.
We apply the ansatz illustrated in Fig.~\ref{qulacs_new_Ansatz}, already used for the SSH model, to the Kitaev chain. The method proves very robust for this second model as well, yielding a fidelity very close to $1$ between the VQE state and the exact state, across the entire range of the $\mu/t$ ratio. 
This also allows us to determine the location of the topological quantum phase transition point, without assuming a priori knowledge about its location. 
The purities between nearest-neighbor sites in the two states also match completely.
Similar to the SSH simulations, we used $\lceil\log_2(n)\rceil + 1$ repetitions of layers, which indicates good scalability of the method with the number of sites. This scaling has been shown to maintain high fidelity even as the number of qubits increases, as demonstrated by simulations performed for systems up to $14$ qubits in the topologically non-trivial phases of both the SSH, for representative $\delta = 0.6$, and Kitaev, for representative  $\mu/t = 1.0$, models. No significant drops in fidelity were observed as the system size increased, as is shown in Fig.~\ref{scaling}, confirming the robustness of the chosen scaling approach. The fidelity remained exceptionally high ( $> 0.999$ ) across all tested system sizes, with a gradual and regular decrease attributed to the increasing complexity of the ground state representation as the system size grows. Testing larger systems, however, requires significantly more computational resources, both for executing the VQE algorithm and for computing exact reference ground states. Nonetheless, our simulations up to $14$ qubits demonstrate that the scaling provides good performance, ensuring high fidelity as the system size increases moderately, thus hinting towards the scalability of the approach.

\begin{figure}[t]
\centering
\includegraphics[width=0.49\textwidth]{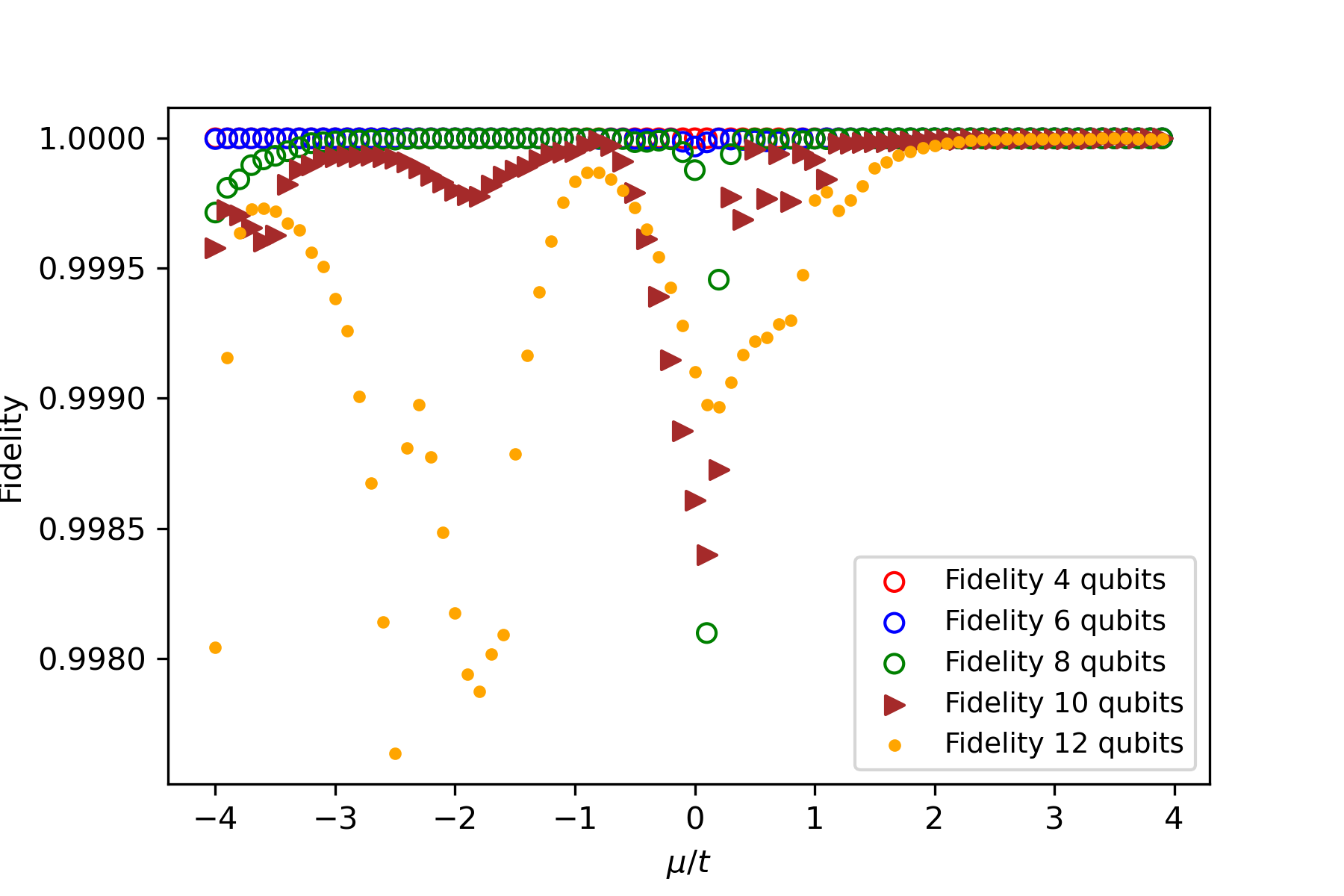}
\caption{Results for the Kitaev chain, exploiting $N=4,6,8,10,12$ sites, and assuming $t=1$ and $\Delta = 1.2$. The fidelities between the exact and the VQE states are shown, demonstrating good agreement across the entire reported range of $\mu/t$.}
\label{kitaev-statevec-12q}
\end{figure}

The comparison between $N= 4, 6, 8, 10$, and 12 sites is shown in Fig.~\ref{kitaev-statevec-12q}, where an excellent convergence of the algorithm to the exact ground state is preserved, also with increasing system sizes.

Further results obtained following separately the second and third strategies mentioned in section \ref{VQEsim} are reported in the appendices \ref{adaptapp} and \ref{warmapp}.

\section{Conclusions} \label{concl}

In this work, we applied the VQE algorithm on two one-dimensional fermionic models exhibiting non-trivial (symmetry-protected) topological phases, namely the SSH and Kitaev chains. Using a hardware-efficient ansatz, we found that the VQE algorithm fails to determine the ground state in the topologically non-trivial phases of both models, due to its (quasi-)degeneracy.
Notably, due to topology, this quasi-degeneracy cannot be removed by small perturbations, differently as, for instance, in typical spin systems. This fact makes the considered chains even more suitable to illustrate the power of our method.

Different methods for improving the results of the standard hardware-efficient ansatz have been investigated, taking into account the topological properties of the considered models. In particular, we added constraints to the cost function related to entanglement witnesses between two qubits, controlling the parameters of the initial ansatz state, and performed adiabatic assistance, using previously obtained solutions for the next VQE initial state. Although in this way some improvement can be achieved, the best results have been obtained by a problem-inspired ansatz, which incorporates some features of the considered models. Interestingly, the problem-inspired ansatz may assure good scalability, as it achieves near-unit fidelities with up to 14 qubits, and it does not require any a priori knowledge of the phase diagram of the system.

The problem-inspired ansatz, utilized in this investigation, is composed of single qubit $R_y$ and two-qubit $R_N$~\eqref{eq:number_preserving} gates between nearest-neighbor qubits, mimicking the Hamiltonian terms of the considered models, and adding an additional set of the latter between the first and last qubit to generate the correlations present in the topologically non-trivial phase. By utilizing only the average energy as a cost function, the VQE is able to determine the correct ground state both in the topologically trivial and non-trivial phase. As a consequence, this enables to determine the topological quantum phase transition point by evaluating, on the obtained state, known properties of the two phases, e.g., the edge-state concurrence and the nearest-neighbor two-qubit purity for the SSH and the Kitaev model, respectively.

Our investigation reinforces the applicability of VQE algorithms for the determination of a ground state and its energy in cases where the latter is degenerate or quasi-degenerate, as in some topologically non-trivial phases. Indeed, it proved able to resolve efficiently the degeneracy and also to determine precisely the topological phase transition point.
More in detail, our approaches to drive the VQE algorithm by specific properties proved particularly effective.

Various system with topologically-protected  degeneracy are known in literature or can be envisaged, generally not allowing direct mapping on free fermionic states. 
Similar as in our work, systems with symmetry-protected topological order can host 
protected ground-state degeneracies when protected modes are present.
Examples are two dimensional p-wave superconductors with half-vortices.
This mechanism works also for systems with genuine topological order, as non-Abelian fractional quantum Hall systems. like Moore-Read states.
There, one can have degeneracies trapping anyons (low-energy excitations), for instance with local potentials.
For reference in this set, see \cite{RevModPhys.80.1083}.
Degeneracies from anyons can be obtained also by proximity effect, as from a Laughlin state and a superconductor or a ferromagnets \cite{stern2012,clarke2013}. 
Using three-dimensional models, it is possible to build Majorana edge modes at the interface between a topological insulator and an adequate number of superconductors, along the same lines as in \cite{fu2008}.
Another relevant and wide set is represented again by two-dimensional systems with topological order, when on not-trivial manifolds and/or with not-trivial boundaries.
In this set, we find toric codes on the torus, surface codes with suitable boundaries (2 rough and 2 smooth), surface codes with holes, FQH states on the torus \cite{kitaev1998,Kitaev_2003}. 
Finally, surface codes also allow the possibility to exploit one-dimensional dislocations \cite{bombin2010,wen2012}.

Further central developments involve the addition and consequent mitigation of realistic or simulated noises, inducing quantum decoherence. As mentioned in the Introduction, this problem has been already shown critical for all the existing quantum architectures.\\

{\bf Acknowledgements --} 
L. L. is pleased to thank Michele Burrello, Stefano Carretta, and Paolo Santini for support and discussions.\\
L. L. also acknowledges financial support by a project funded under the National Recovery and Resilience Plan (NRRP), Mission 4 Component 2 Investment 1.3 - Call for tender No. 341 of 15/03/2022 of Italian Ministry of University and Research funded by the European Union – NextGenerationEU, award number PE0000023, Concession Decree No. 1564 of 11/10/2022 adopted by the Italian Ministry of University and Research, CUP D93C22000940001, Project title ``National Quantum Science and Technology Institute'' (NQSTI).\\
S. P. acknowledges financial support from the University of L'Aquila by the internal project ``Variational Quantum Eigensolver methods for the disordered Su-Schrieffer-Heeger model''and the Ministero dell'Università e della Ricerca (MUR) and the Project PRIN 2022 number 2022W9W423 funded by the European Union Next Generation EU.\\
We acknowledge the CINECA award under the ISCRA initiative, for the availability of high-performance computing resources and support.\\
M.C. acknowledges financial support by the Quantum Thermodynamics of Precision in Electronic Devices (ASPECTS) project from the Horizon Europe funding agency --- HORIZON-WIDERA-2022-ACCESS-07 grant agreement 101080167. T.J.G.A. acknowledges financial support by Xjenza Malta, through project APTLY-Quantum topological system tackled by quantum computations (IPAS-2023-026).\\
M. B. acknowledges computational resources by Variational Quantum Eigensolver for Topological systems (OPEN-30-32) project from National Supercomputing Center IT4Innovations.\\ The views and opinions expressed are those of the author(s) only and do not necessarily reflect those of the European Union. Neither the European Union nor the granting authority can be held responsible for them.

\appendix

\section{Concurrence of a two-qubit state and Purity }
\label{appconc} 

In this appendix, we define two quantities that we use here to quantify the entanglement between two parts of the analyzed system: the concurrence and the purity. 

Given a two-qubit state represented by a density-matrix $\rho$, the concurrence is an entanglement monotone defined as \cite{CONC1Hill_1997, CONC2PhysRevLett.80.2245, CONC3hildebrand2007concurrence, CONC4RevModPhys.81.865}:
\begin{equation}
\mathit{C}(\rho) \equiv \text{max}(0, \lambda_1 - \lambda_2 - \lambda_3 - \lambda_4 ),
\end{equation}
where $\lambda_1,...,\lambda_4$ are the eigenvalues, in decreasing order, of the Hermitian matrix 
\begin{equation}
R=\sqrt{\sqrt{\rho}\tilde{\rho} \sqrt{\rho}} ,
\end{equation}
with $\tilde{\rho}$ defined by 
\begin{equation}
\tilde{\rho}=(\sigma_y \otimes \sigma_y)\rho^*(\sigma_y \otimes \sigma_y),
\end{equation}
and $\sigma_y$ is the Pauli matrix. The concurrence ranges from $0$ for a separable state to $1$ for a maximally entangled state.

The purity is defined for a generic state $\rho$ as
\begin{equation}
\mathcal{P}\left(\rho \right) = \mathrm{Tr} \left(\rho^2 \right) \, ,
\end{equation}
and it is bounded from above by $1$ for pure states, and from below by $1/d$ (where $d$ denotes the dimension of the Hilbert space) when the state is maximally mixed. In quantum computation, it is often useful to assess the purity of quantum states, particularly in the context of quantum error correction and fault-tolerant quantum computing~\cite{Xu2023}. Given a pure state, the purity of the reduced density matrix of a subsystem is an indicator of how much it is entangled with the rest of the system.

\section{Adapted cost function}
\label{adaptapp}
In this appendix, we described in more detail the second strategy mentioned in section \ref{VQEsim} on the main text, and we report some results obtained following it.

In the SSH model, the ground state in the non-trivial phase is expected to exhibit a significant entanglement between the edge sites. Moreover, because of the dimerization effect, the bulk tends to be arranged in singlets of adjacent sites. Since the two phases, topological and not, are characterized by two different dimerizations, two adjacent sites will be highly entangled in one phase and almost disentangled in the other. These are the features of the SSH model that we exploit here. The amount of entanglement between the edge sites can be quantified by the concurrence of the corresponding reduced density matrix. Moreover, the different bulk dimerizations give rise to an alternation of values of the purity between pairs of adjacent sites (see the appendix~\ref{appconc} for definitions). We know that the concurrence between the edges is very small in the trivial phase (vanishing in the thermodynamic limit), while it increases rapidly when going deeper into the topologically non-trivial phase.

\begin{figure}[H]
\centering
\includegraphics[width=0.51\textwidth]{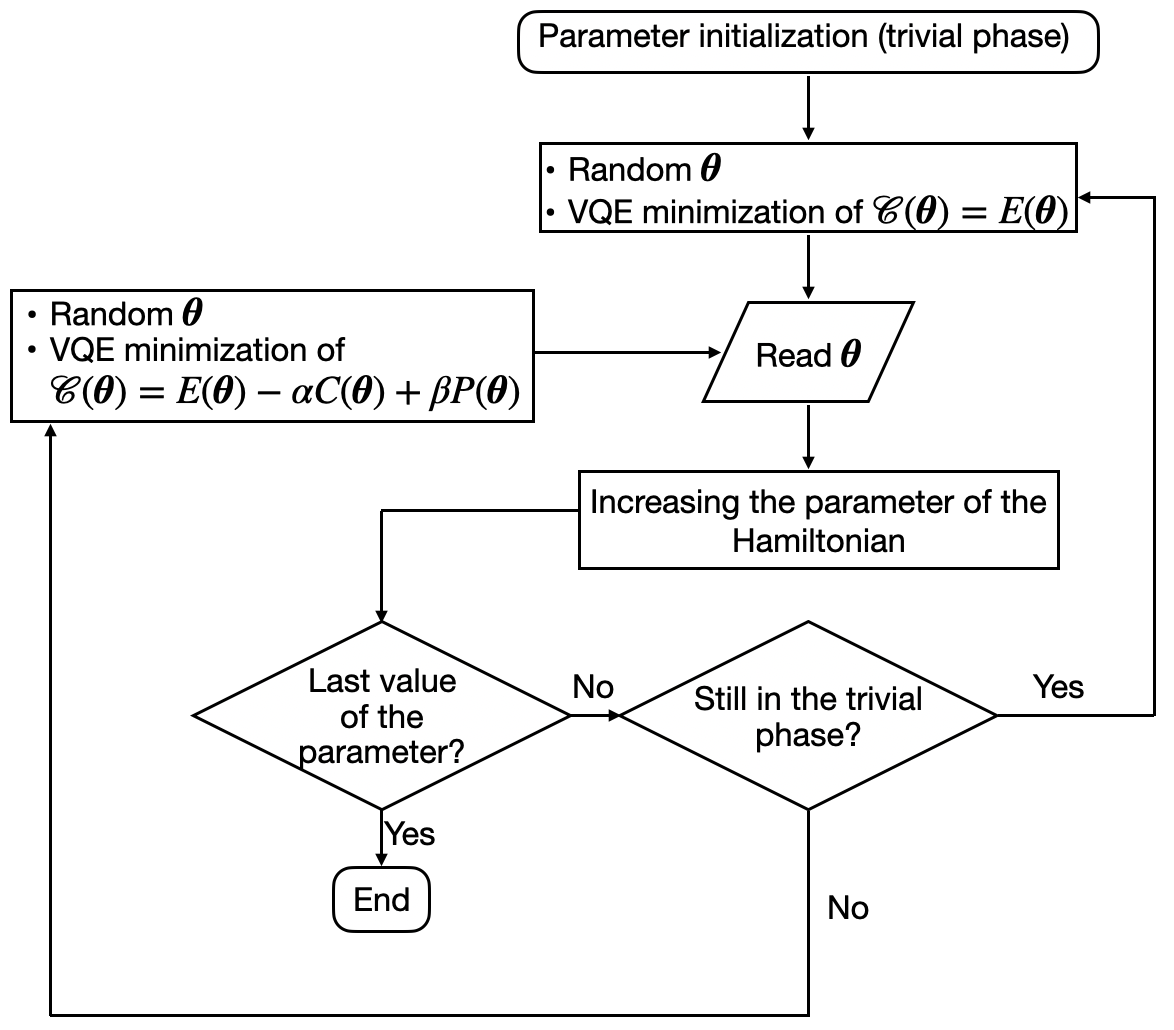}
\caption{Scheme of the modified cost function approach to the VQE.}
\label{diag1}
\end{figure}

In the topologically non-trivial phase, the purity of the reduced density matrix for sites connected by the hopping $v$ switches from values close to $1$,
in the trivial phase, to values close to $1/4$, in the non-trivial phase. Hence, additional constraints can be inserted into the cost function, pushing it to search for the minimum energy state and, at the same time, minimizing the purity between pairs of nearest-neighboring bulk sites with hopping amplitude $v$, and maximizing the concurrence between the edges. These tasks are performed according to the following equation
\begin{equation}
\mathcal{C}(\bm{\theta})=E(\bm{\theta}) - \eta \, C\left[\rho_\text{edge}(\bm{\theta})\right] + \tau \, \mathcal{P}\left[\rho_\text{nn}(\bm{\theta})\right],
\label{cost_purity}
\end{equation} 
where where $\bm{\theta}$ denotes the vector of the ansatz parameters, $E(\bm{\theta})$ is the average energy, $\rho_\text{edge}$ is the reduced density matrix of the edge sites, $\rho_\text{nn}$ is the reduced matrix of two nearest-neighbor sites, $\eta$ and $\tau$ are two Lagrangian coefficients. 
In this approach, the initial minimization step is performed with respect to the parameters $\bm{\theta}$, while keeping the coefficient $\eta$ constant, so as to approach the almost-degenerate subspace. The minimization is then restarted with the final parameters, gradually increasing $\eta$ at each repetition, until a sufficient minimum of the cost function is achieved.
As a result, the fidelity between the VQE state and the exact state is increased beyond $0.99$. The scheme is depicted in Fig. \ref{diag1}

\begin{figure}[H]
\centering
\includegraphics[width=0.49\textwidth]{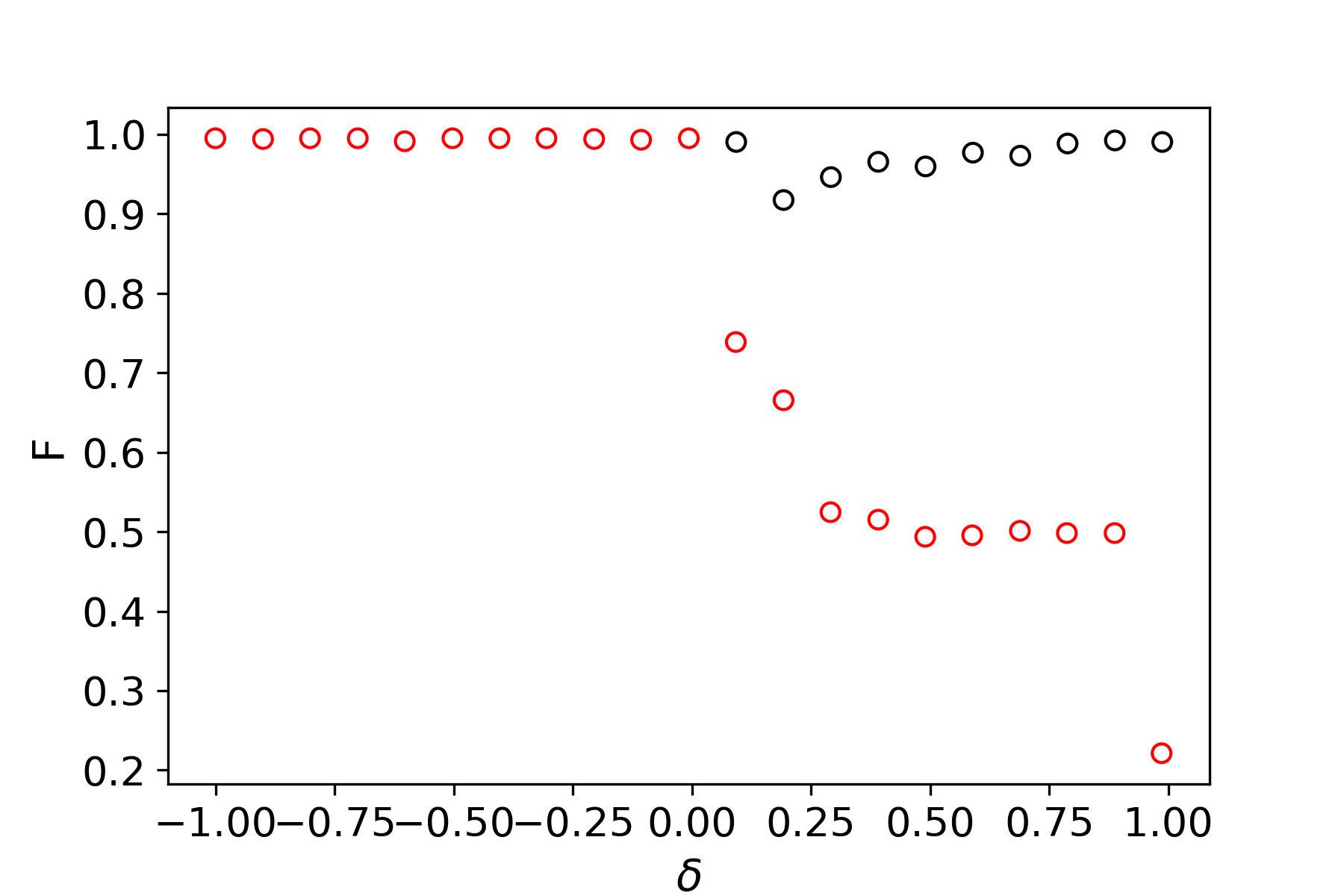}
\caption{Fidelity between the VQE state and exact ground state of the SSH chain, as a function of $\delta$ and for $N=6$ sites. The cases are reported when no constraints are encoded in the cost function (red circles) and when the concurrence and purity constraints are imposed (black circles). In the latter situation, we achieved convergence in the topologically non-trivial phase, maintaining the fidelity values ranging from $0.92$ to over $0.99$.}
\label{fidelity_ssh}
\end{figure}

In Fig.~\ref{fidelity_ssh}, the fidelity between the VQE state and the exact ground state is shown, as a function of the $\delta$ parameter for $N=6$ sites and \emph{a}) when the cost function contains the energy term only; \emph{b}) when the concurrence and purity penalty terms are also included. 

\begin{figure}[H]
\centering
\includegraphics[width=0.40\textwidth]{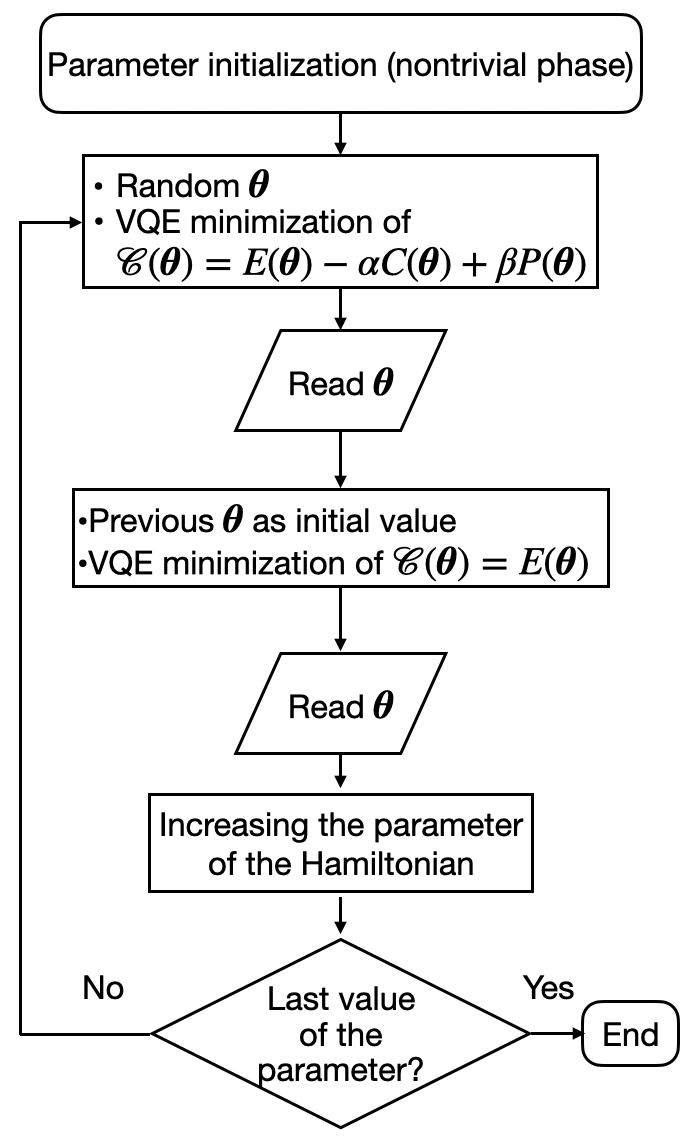}
\caption{Scheme of the modified cost function approach to the VQE enhanced by a warm-start passage.}
\label{diag2}
\end{figure}

A Warm-Start approach has been also employed to further improve the convergence beyond the values already obtained, without using constraints on the cost function \cite{egger2021}. This technique is useful to speed up the convergence of the algorithm, by initializing the optimization procedure with parameters obtained from previous runs. Here, we initialize the state with the one obtained imposing the extra constraints on the purity and on the concurrence, and then we run the VQE minimizing only the mean energy, as depicted in Fig. \ref{diag2}.

\begin{figure}[H]
\centering
\includegraphics[width=0.49\textwidth]{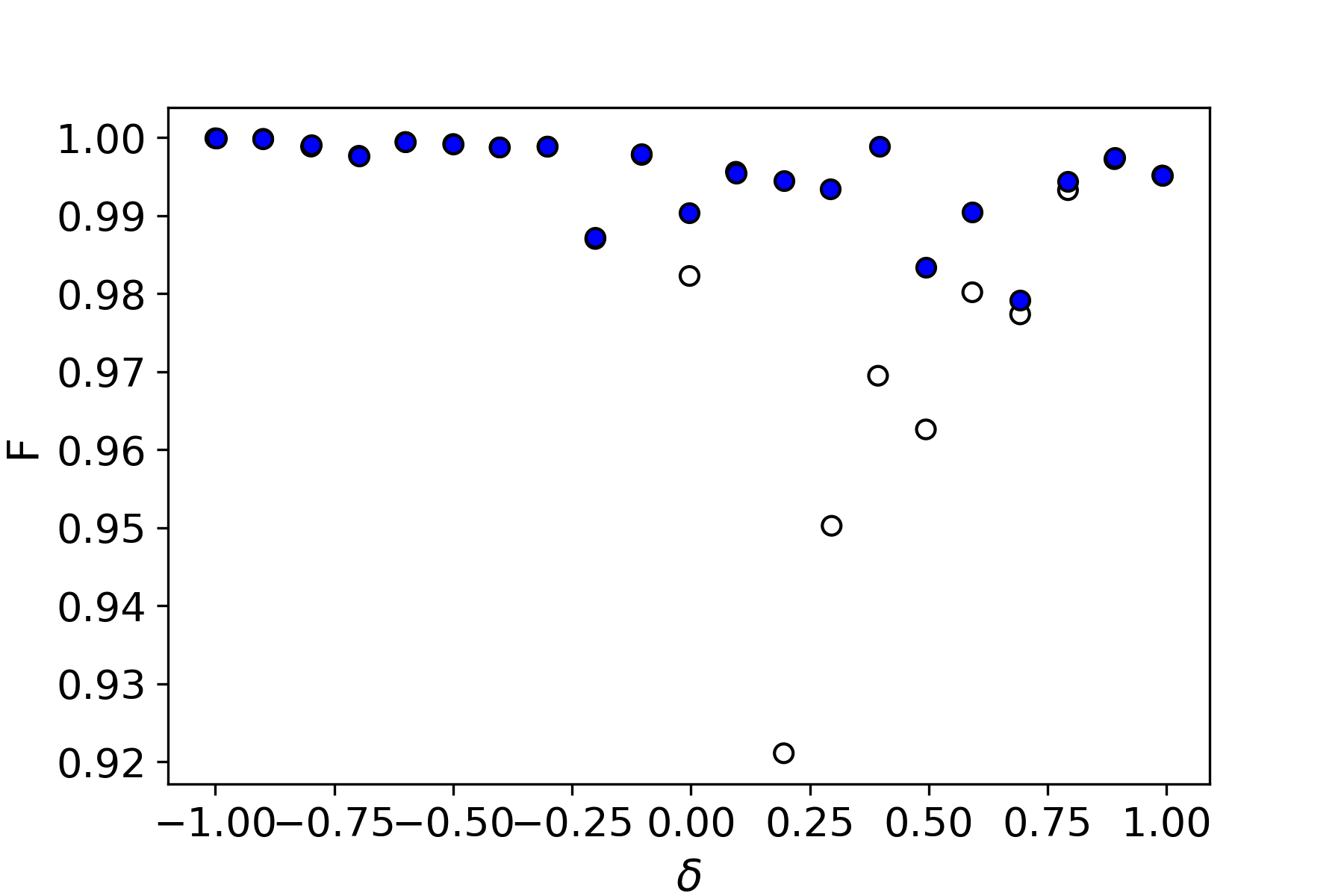}
\caption{Comparisons between the fidelity of the VQE state and the exact ground state of the SSH chain with $N=6$ sites, obtained imposing the extra constraints on the cost function (black circles) and by the Warm-Start approach (blue dots).}
\label{warm_start_SSH}
\end{figure}

The results, as functions of $\delta$, are shown in Fig.~\ref{warm_start_SSH} for $N=6$ sites.

\begin{figure}[H]\label{improvements}
\centering
\includegraphics[width=0.49\textwidth]{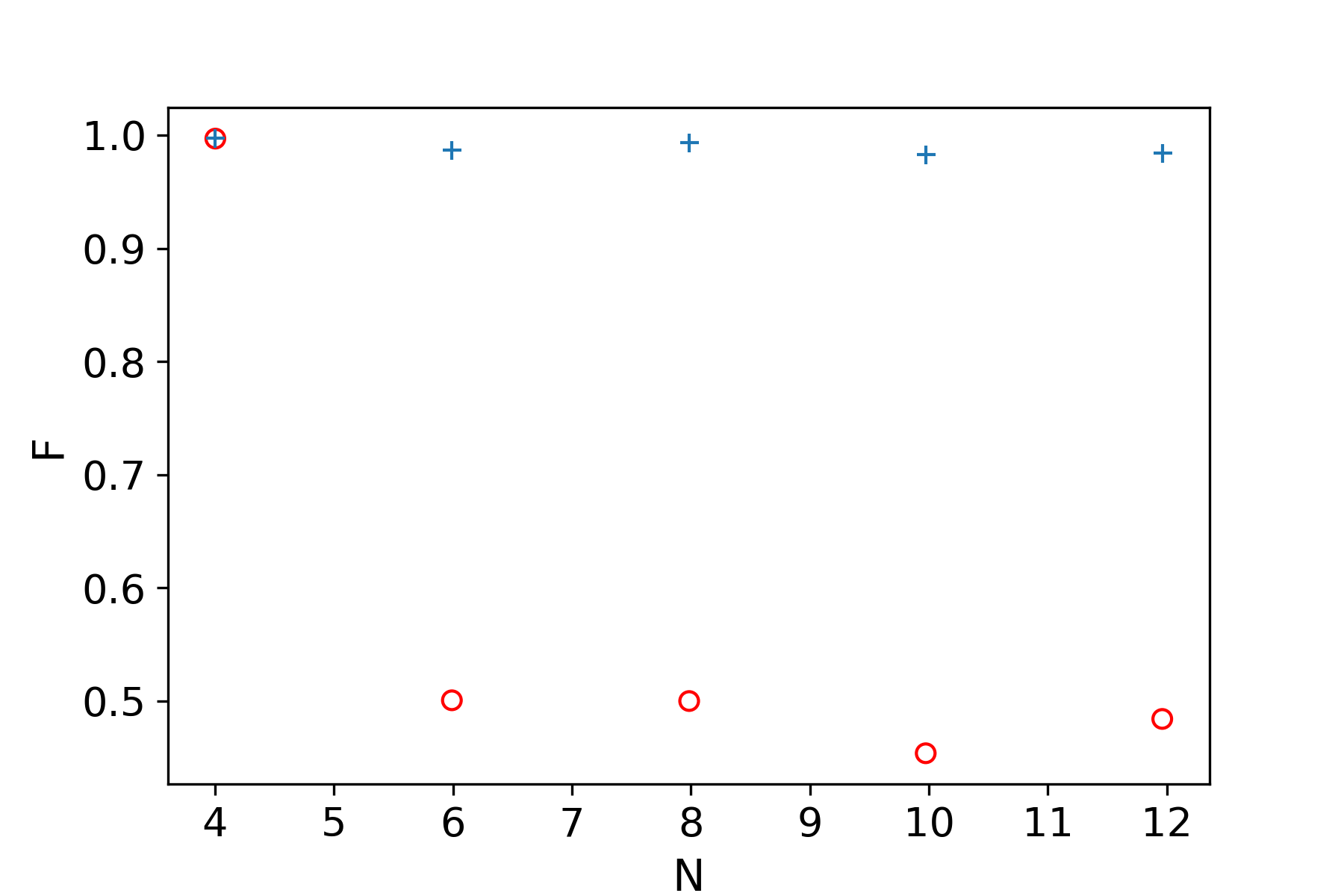}
\caption{Fidelity between the VQE state and the exact ground state of the SSH chain, as a function of the number of sites $N$, for the unconstrained VQE (red circles) and for the Warm-Start approach (blue crosses), for $\delta=0.8$. It is noteworthy how the cost function with the constraints from the concurrence and from the purity, as well as with the addition of the Warm-Start approach, ensures VQE convergence, regardless of the assumed number of sites, up to $N = 12$.}
\label{Fidelity_Nvar}
\end{figure}

Moreover, in Fig.~\ref{Fidelity_Nvar}, the fidelity between the VQE state and the exact ground state is depicted as a function of the number $N$ of sites and at $\delta = 0.8$, for both the VQE without any constraints in the cost function, and with the Warm-Start approach. It can be observed that, in the latter situation, there is a significant improvement in the accuracy of the algorithm.

As found for the SSH model, the fidelity in the Kitaev chain between the exact ground state and the corresponding VQE state decreases relevantly in the non-trivial phase, as shown in Fig.~\ref{fidkit} (red circles). 
Similarly, if we consider the reduced density matrix between two adjacent sites of the Kitaev chain, its purity has a sudden decrease in the non-trivial phase. This fact can be intuitively understood by observing that the XY spin model (where the Kitaev chain is mapped on by the Jordan-Wigner transformation) exhibits a ferromagnetic phase when the Kitaev model is in its non-trivial phase. Here the ground state is a GHZ entangled state (due to the $Z_2$ spin-flip symmetry of the XY Hamiltonian \cite{PhysRevA.88.052305,mussardo}), then every two-spin state is maximally mixed.

\begin{figure}[H]
\centering
\includegraphics[width=0.49\textwidth]{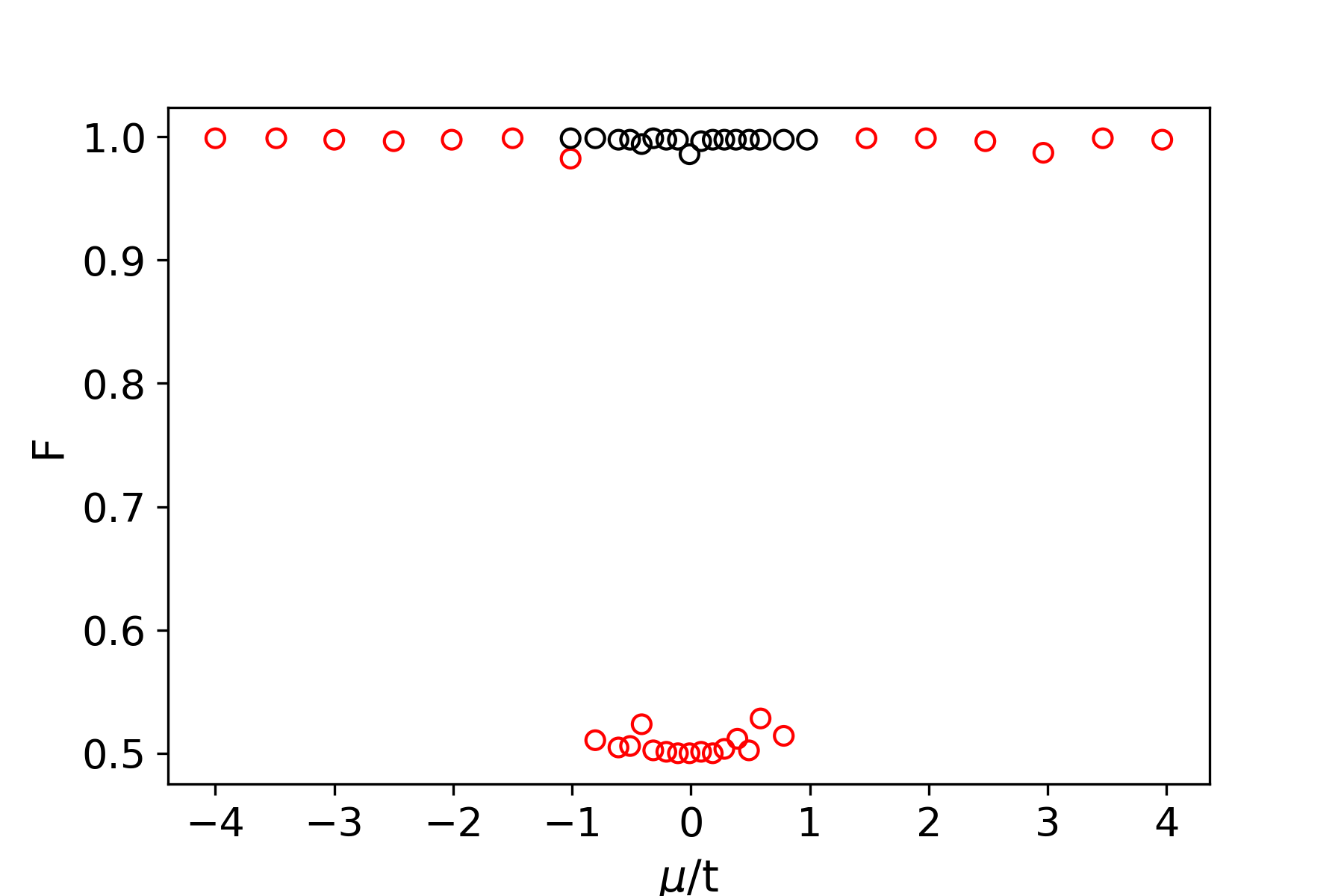}
\caption{Fidelity between the exact ground state of the Kitaev chain and the corresponding VQE states with constraints (red circles) and without constraints (black circles), as functions of $\mu/t$, $\Delta/t =1.2$, and for $N=6$ sites.}
\label{fidelity_kit}
\end{figure}

\begin{figure}[H]
\centering
\includegraphics[width=0.45\textwidth]{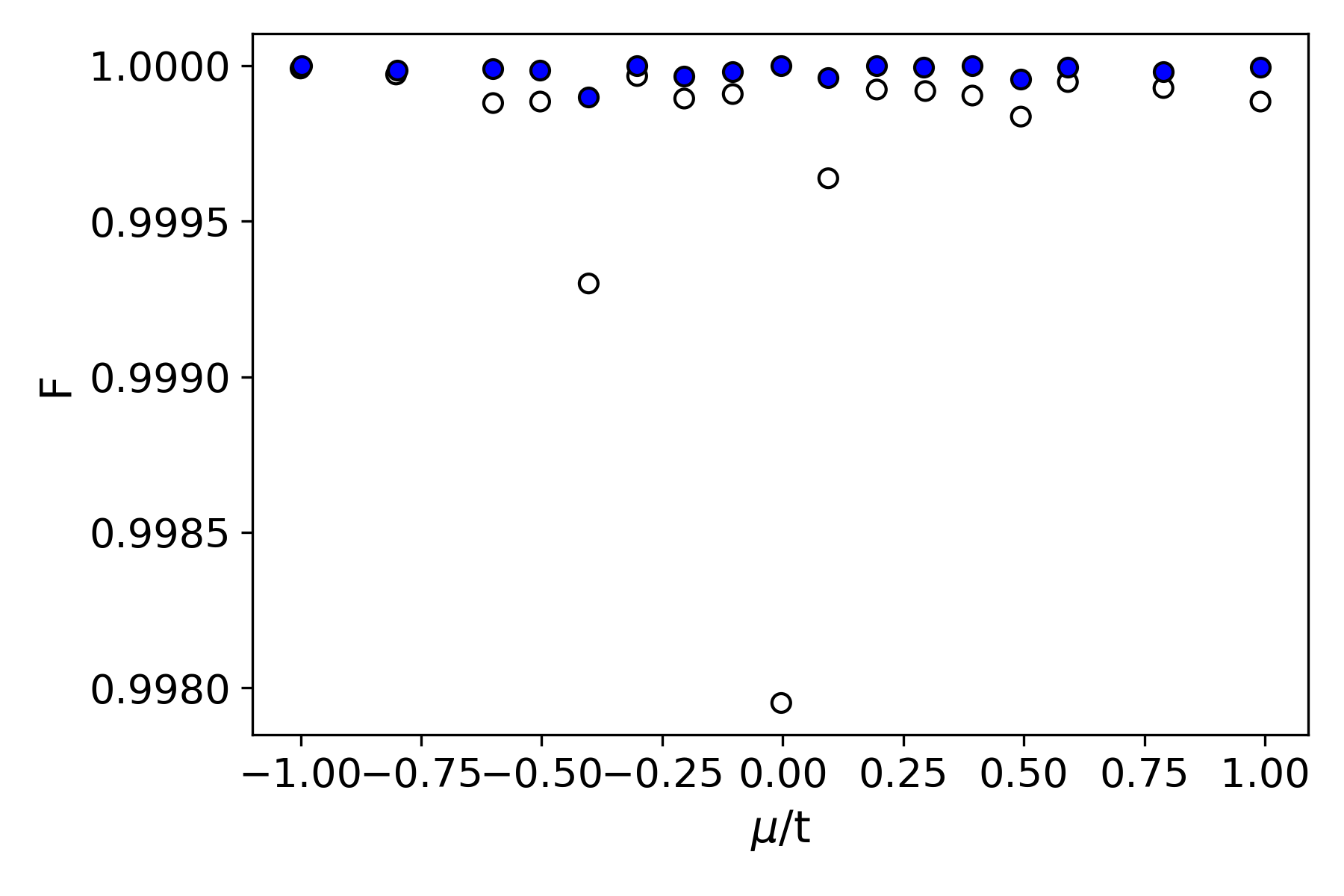}
\caption{Fidelity between the VQE state and the exact ground state of the Kitaev chain with $N=6$ sites, $\Delta/t = 1.2$, and as a function of the ratio $\mu / t$. The black circles represent the data obtained from the VQE runs with constraints, while the blue dots represent the data collected from the Warm-Start approach.}
\label{WSkit}
\end{figure}

By introducing the constraint of minimizing the purity, in the range for the ratio $\mu/t$ where the algorithm without any constraints fails, we achieve a significant improvement in the fidelity, as shown in Fig.~\ref{fidelity_kit} (black dots). 

Again, we also exploit the Warm-Start approach to further enhance the accuracy already achieved in the topologically non-trivial phase with the purity as a constraint. This leads to a significant improvement of the fidelity, as shown in Fig.~\ref{WSkit}.

\begin{figure}[H]
\centering
\includegraphics[width=0.49\textwidth]{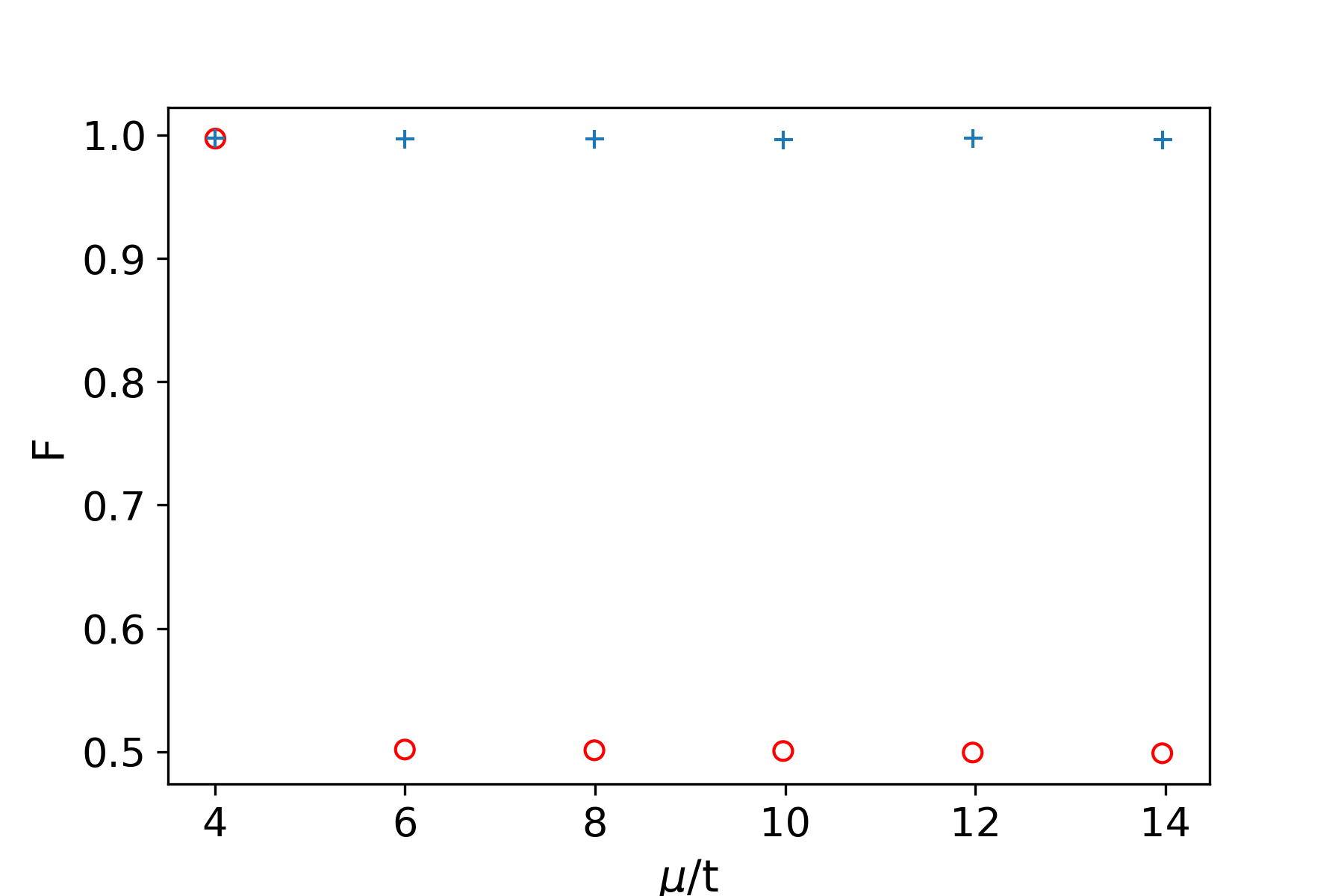}
\caption{Fidelity between the VQE state and the exact ground state of the Kitaev chain, as a function of the number $N$ of sites, and with fixed values for the ratio $\mu / t = 0.3$ and for $\Delta/t = 1.2$. The blue crosses are the data collected from the Warm-Start approach, while the red circles denote the fidelity values obtained from the VQE runs without constraints in the cost function.}
\label{Nkit}
\end{figure}

In Fig.~\ref{Nkit}, we show the fidelity as a function of the number $N$ of sites, for $\mu/t=0.3$ and $\Delta/t=1.2$, in the case of the VQE without the constraints (red circles) and with the Warm-Start approach (blue crosses), up to $N = 14$ sites (which is the largest number of sites we were able to test on our high-performance computing facilities). 

\section{Progressive change of parameters with concurrence and purity controls} 
\label{warmapp}
In this appendix, we described in more detail the combination of the second and third approaches mentioned in section \ref{VQEsim} on the main text, and we report some results obtained following it.

\begin{figure}[h]
\centering
\includegraphics[width=0.40\textwidth]{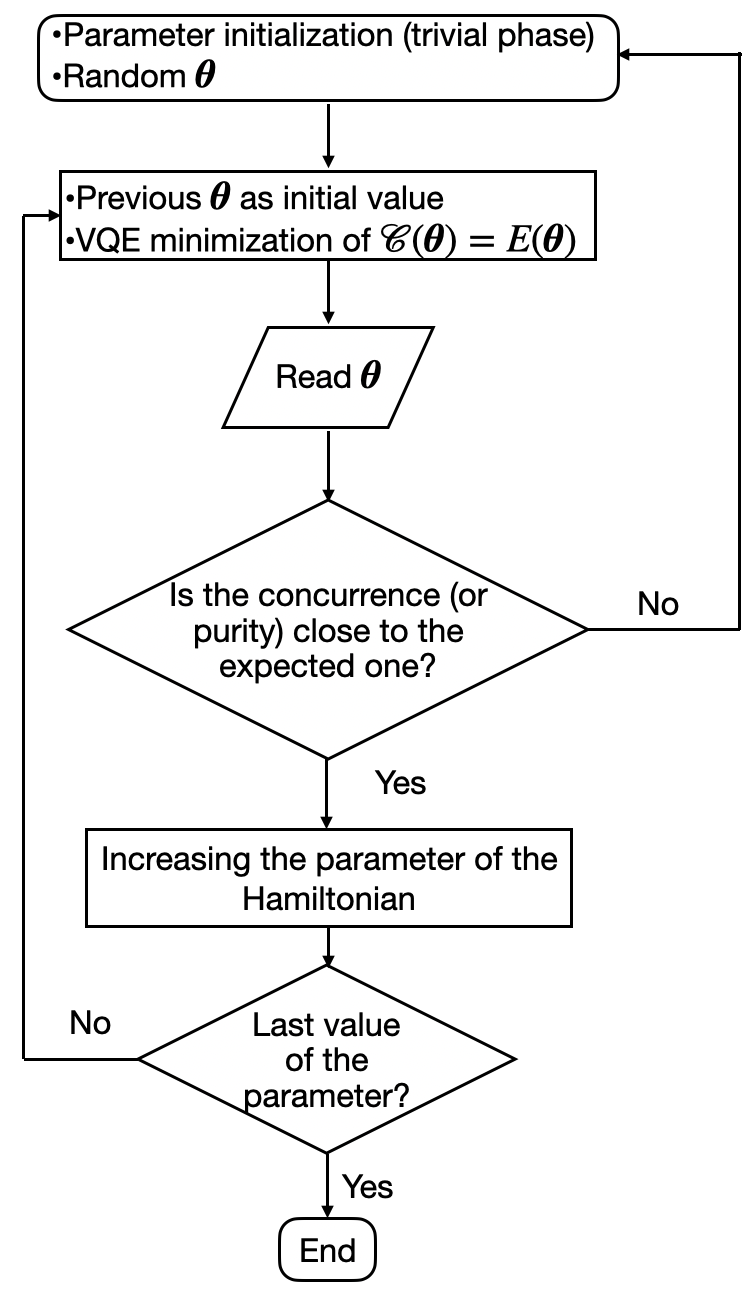}
\caption{ Scheme of the approach based on the slow increment of the parameters and subsequent check of the concurrence and/or purity.}
\label{diag3}
\end{figure}

Concerning the random choice of the initial parameters in the circuit, the problem of choosing the suitable Lagrangian multipliers in Eq.~\eqref{cost_purity} requires a careful investigation. For this reason, here we propose another Warm-Start approach, optimizing only the energy of the system, i.e., defining the cost function
\begin{align}
 \mathcal{C}(\bm{\theta}) = \min_{\bm{\theta}} E\left(\bm{\theta}\right),
\end{align}
but also introducing a result acceptance check, based on the values of the concurrence for the SSH chain and of the purity for the Kitaev chain. 
Starting from $\delta_{\text{first}}=-1.0$ for the SSH chain ($\mu /t$ for the Kitaev chain) in the trivial phase, the first-point algorithm takes the number $n$ of initial guesses as an input, generating directly the pool of competing points from which the one with the lowest energy is taken. Thereafter, one checks if the concurrence of its edge qubits is close to 0 (the purity of two first neighbors is close to 1), and proceeds with the computation of the subsequent points, each time starting from the reference parameters of the previous point. This scheme is depicted in Fig. \ref{diag3}.

\begin{figure}[h]
\centering
\includegraphics[width=0.49\textwidth]{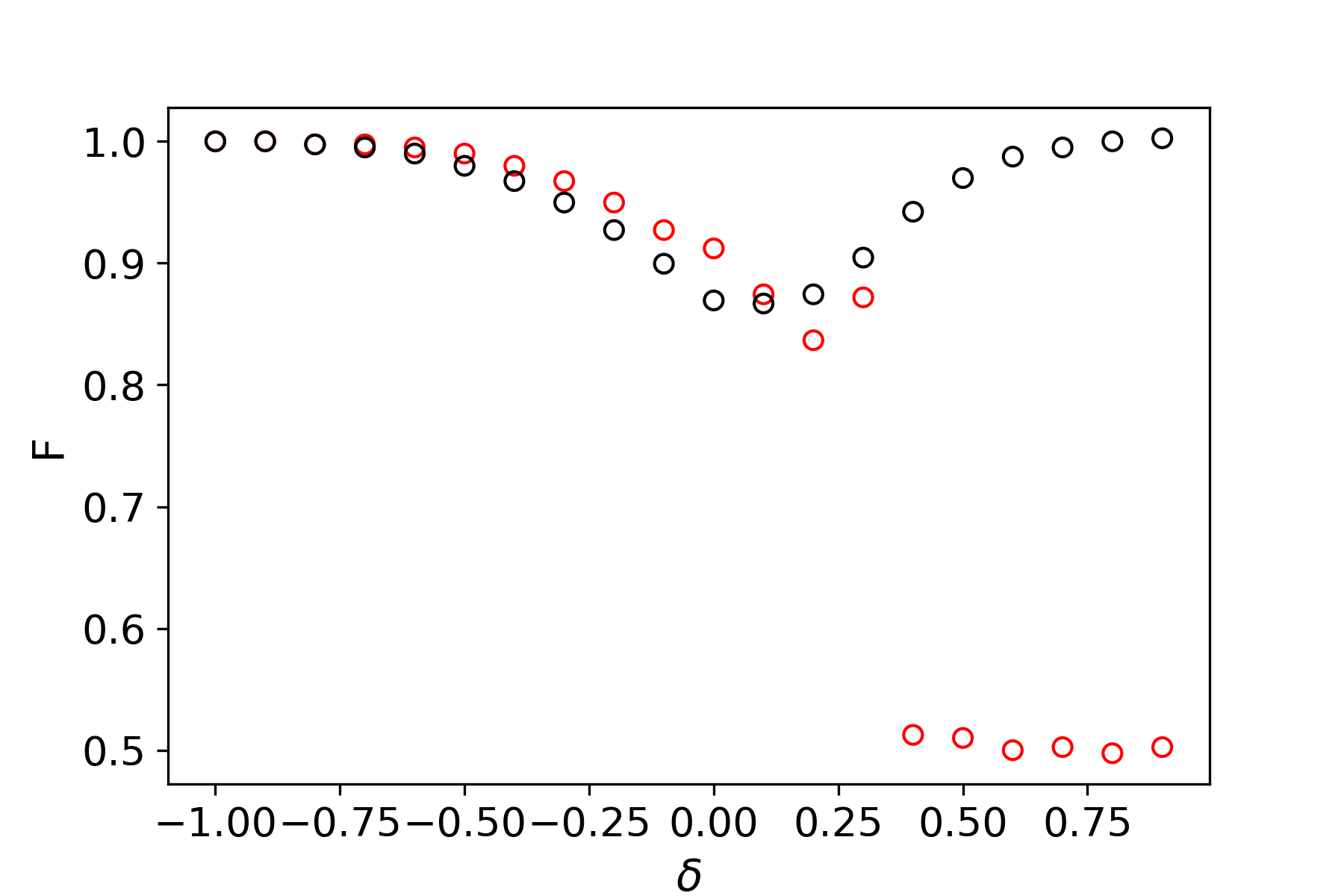}
\caption{Results for the SSH chain with $N=10$ sites. The black circles represent the fidelity values between the VQE state obtained imposing the concurrence control and the state from exact diagonalization, while the red circles represent the fidelity between the VQE state derived without any concurrence control and the exact state.}
\label{Fid_check_C}
\end{figure}

The results for the SSH model with $N = 10$ sites are shown in the Fig.~\ref{Fid_check_C}, where the black circles denote the fidelity between the VQE state obtained using the present method and the exact state, while the red circles represent the same fidelity but without performing any concurrence check. A minimum fidelity around $0.87$ is observed at the transition point $\delta=0$, increasing beyond $0.99$ in the topologically non-trivial phase, as the fully dimerized limit is approached. Instead, the VQE simulations without the concurrence control return a minimum-energy state that is a linear superposition of the two degenerate ground states, with fidelity dropping down to around 50\% in the topologically non-trivial phase.

\begin{figure}[t]
\centering
\includegraphics[width=0.49\textwidth]{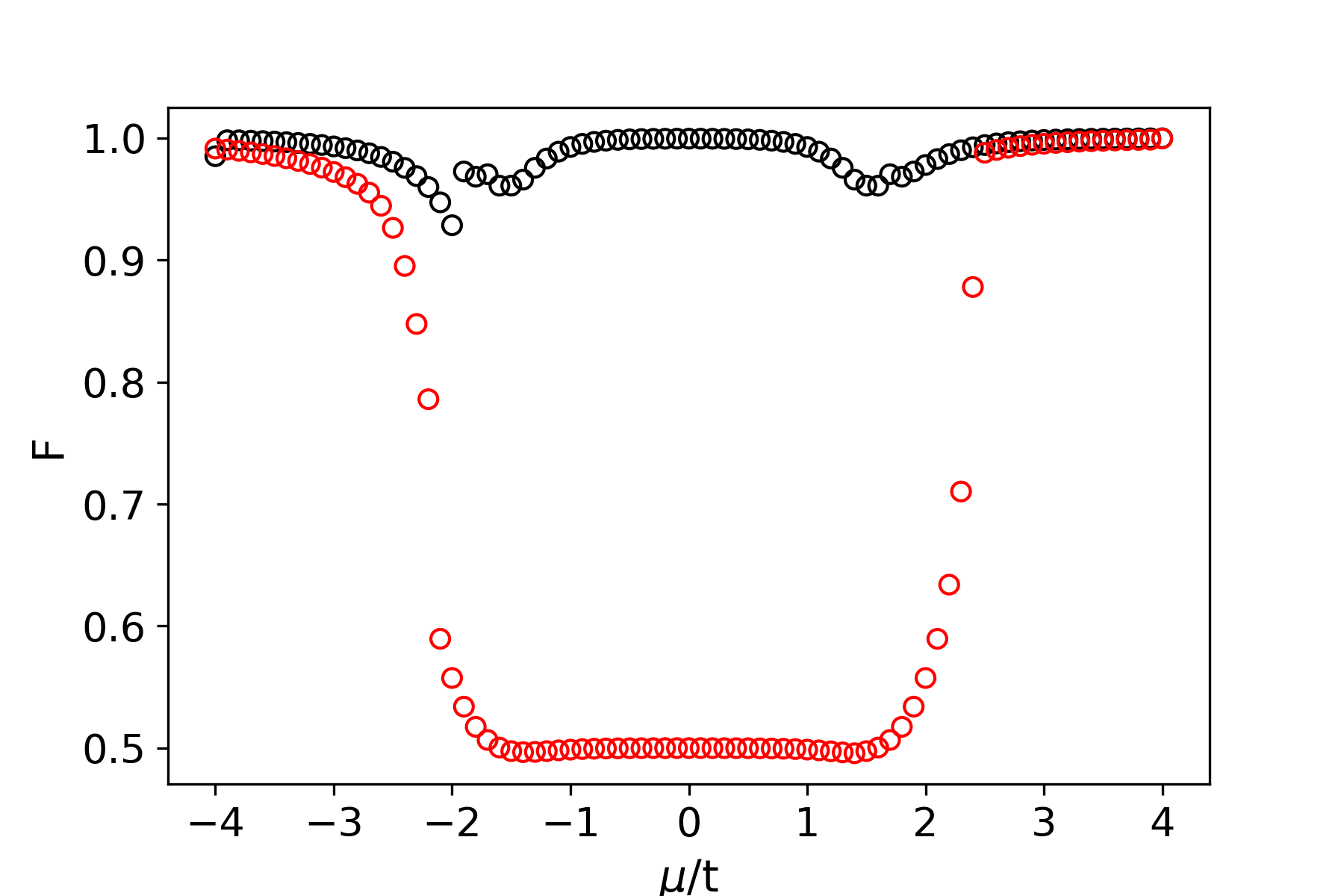}
\caption{Results for the Kitaev chain with $N=10$ sites, assuming $t=1$ and $\Delta = 1.2$. The black circles represent the fidelity values between the VQE state obtained from the simulation with the purity control and the state from exact diagonalization, while the red circles represent the fidelity between the VQE state obtained without any purity control and the exact state.}
\label{Fidelities_10q}
\end{figure}

In Fig.~\ref{Fidelities_10q}, the results for the Kitaev chain, with $N=10$ sites, $t=1$, and $\Delta = 1.2$ are presented. The black circles show the fidelity, as the ratio $\mu/t$ varies, of the exact state with the VQE state obtained using the purity control method, while the red circles show the same fidelity with the VQE state obtained without the purity control. This method improves the convergence of the algorithm to over 99\% in the topologically non-trivial phase, with minimum values of 92\% and 96\% around the phase transition points. These results are in contrast to the fidelity values obtained by the VQE algorithm without purity check, where the fidelity drastically drops to 50\%.

\clearpage

\bibliography{bbb}

\end{document}